\documentclass[fleqn,usenatbib]{mnras}

\usepackage{newtxtext,newtxmath}

\usepackage[T1]{fontenc}

\DeclareRobustCommand{\VAN}[3]{#2}
\let\VANthebibliography\thebibliography
\def\thebibliography{\DeclareRobustCommand{\VAN}[3]{##3}\VANthebibliography}

\usepackage{graphicx}	
\usepackage{amsmath}	
\usepackage{subfigure}
\usepackage{scalerel}

\newcommand\HII{H\protect\scaleto{$II$}{1.2ex}}
\newcommand\CIV{C\protect\scaleto{$IV$}{1.2ex}}
\newcommand\OII{O\protect\scaleto{$II$}{1.2ex}}
\newcommand\lya{Ly$\alpha$}
\newcommand\ha{H$\alpha$}
\newcommand\SigmaM{$\rm \Sigma_{M_\star}$}
\newcommand\SigmaSFR{$\rm \Sigma_{SFR}$}

\graphicspath{{figures/}} 

\title[Properties of stellar clumps in redshift 4-to-5 galaxies]{Properties of the brightest young stellar clumps in extremely lensed galaxies at redshifts 4 to 5}

\author[M. Messa et al.]{
Matteo Messa,$^{1,2,3}$\thanks{E-mail: matteo.messa@inaf.it}
Miroslava Dessauges-Zavadsky,$^{2}$
Angela Adamo,$^{1}$ 
Johan Richard$^{4}$
\newauthor
and Adélaïde Claeyssens$^{1}$
\\
$^{1}$The Oskar Klein Centre, Department of Astronomy, Stockholm University, AlbaNova, SE-10691 Stockholm, Sweden\\
$^{2}$Observatoire de Genève, Université de Genève, Versoix, Switzerland\\
$^{3}$Osservatorio di Astrofisica e Scienza dello Spazio di Bologna, INAF, Bologna, 40129, Italy.\\
$^{4}$Univ Lyon, Univ Lyon1, ENS de Lyon, CNRS, Centre de Recherche Astrophysique de Lyon UMR5574, Saint-Genis-Laval, France\\
}

\date{Accepted XXX. Received YYY; in original form ZZZ}

\pubyear{2024}

\begin{document}
\label{firstpage}
\pagerange{\pageref{firstpage}--\pageref{lastpage}}
\maketitle

\begin{abstract}
We study the populations of stellar clumps in three high-redshift galaxies, at z=4.92, 4.88 and 4.03, gravitationally lensed by the foreground galaxy clusters MS1358, RCS0224 and MACS0940, respectively. The lensed galaxies consist of multiple counter-images with large magnifications, mostly above $\rm \mu>5$ and in some cases reaching $\rm \mu>20$. We use rest-frame UV observations from the HST to extract and analyse their clump populations, counting 10, 3 and 11 unique sources, respectively. Most of the clumps have derived effective radii  in the range $\rm R_{eff}=10-100$ pc, with the smallest one down to 6 pc, i.e. consistent with the sizes of individual stellar clusters. Their UV magnitudes correspond to $\rm SFR_{UV}$ mostly in the range $\rm 0.1-1\ M_\odot yr^{-1}$; the most extreme ones, reaching $\rm SFR_{UV}=5\ M_\odot yr^{-1}$ are among the UV-brightest compact ($\rm R_{eff}<100$ pc) star-forming regions observed at any redshift. Clump masses span a broad range, from $10^6$ to $\rm 10^9\ M_\odot$; stellar mass surface densities are comparable, and in many cases larger, than the ones of local stellar clusters, while being typically 10 times larger in size. 
By compiling published properties of clump populations at similar spatial resolution between redshift 0 and 5, we find a tentative  evolution of $\rm \Sigma_{SFR}$ and $\rm \Sigma_{M_\star}$ with redshift, especially when very compact clumps ($\rm R_{eff}\leqslant20$ pc) are considered. We suggest that these trends with redshift reflect the changes in the host galaxy environments where clumps form. Comparisons with the local universe clumps/star clusters shows that, although rare, conditions for elevated clump $\rm \Sigma_{SFR}$ and $\rm \Sigma_{M_\star}$ can be found.
\end{abstract}

\begin{keywords}
gravitational lensing: strong -- galaxies: high-redshift -- galaxies: star formation -- galaxies: star clusters: general 
\end{keywords}



\section{Introduction}
Since the first deep observations with the Hubble space telescope (HST), galaxy morphology was recognised to change from disk-like or elliptical into more irregular appearance at increasing redshifts \citep[e.g.][]{abraham1996,brinchmann1998}. In addition, galaxies around the cosmic noon (z$\sim$1-3) are characterised by the presence of bright stellar clumps dominating their rest-frame ultraviolet (UV) morphology \citep[e.g.][]{cowie1995,vandenbergh1996}.
The James Webb Space Telescope (JWST) is bringing new insight into the properties of high-z galaxies, especially at the epoch of re-ionisation ($\rm z\gtrsim7$); the first results seem to confirm the overall morphological evolution traced by HST at lower redshifts, with galaxies at redshift 7-12 characterised by irregular (yet compact) structures and in a minor part ($\sim20\%$) by interaction/mergers \citep{treu2023}.

One of the current main efforts in the community is understanding the link between clump formation (and evolution) and galaxy growth. Initially observed as large structures, with sizes $\sim1$ kpc and masses $\sim10^8-10^9\ M_\odot$ \citep[e.g.][]{elmegreen2007,forsterschreiber2011a,forsterschreiber2011b,guo2012,soto2017}, stellar clumps are being recently studied in gravitationally-lensed fields, where lensing allows to reach resolutions down to $\sim10$ pc in size and $\sim10^6\ M_\odot$ in mass \citep[e.g.][]{livermore2012,livermore2015,adamo2013,johnson2017,cava2018,vanzella2017a,vanzella2017b,vanzella2019,vanzella2021,vanzella2022,mestric2022,messa2022}, and thus to investigate clump substructures avoiding overestimates driven by poor resolutions \citep[e.g.][]{dessauges2017,meng2020}. The exquisite performance of JWST is also contributing to increase the resolution and depth at which clump samples are studied \citep[e.g.][]{claeyssens2023,vanzella2022_sunrise,vanzella2022_a2744}, and recently proved the possibility of observing the progenitors of local globular clusters \citep{mowla2022,claeyssens2023}.

Most of clumps at redshift $\rm z<3$ may have formed \textit{in-situ} within their host galaxies. This scenario is supported by several observational evidence, such as: (1) the redshift evolution of clumpy galaxies, closely following evolution of the overall star formation rate (SFR) volume density and inconsistent with the evolutionary trends of minor and major mergers \citep{lotz2011,shibuya2016}; (2) the presence of numerous clumpy galaxies (at least up to z$\sim3$) still dominated by disk-like appearance \citep{shibuya2016}, with comparable disk scale-heights in case of either presence or absence of clumps \citep{elmegreen&elmegreen2006,elmegreen2017}; (3) the kinematics of the majority of clumpy star-forming galaxies at cosmic noon being dominated by ordered disk rotation (yet with elevated velocity dispersions, \citealp{forsterschreiber2009,wisnioski2015,rodrigues2016,swinbank2017,turner2017,simons2017,girard2018}). Simulations show that such turbulent high-z disks fragment because of gravitational instability and can form gas clouds that turn into stellar clumps \citep{bournaud2014,tamburello2015,mandelker2017}; (4) the observations of very young clumps (age $\lesssim10$ Myr), inconsistent with an "external" origin \citep[e.g.][]{forsterschreiber2011b,zanella2015};
(5) finally, the stellar mass function of these clumps follows a power-law with slope $-2$ \citep{dessauges&adamo2018}, characteristic of nearby star clusters and \HII\ regions (see e.g. the reviews by \citealp{krumholz2019,adamo2020_review}) and expected for stellar structures formed in a hierarchical manner via turbulence-driven fragmentation \citep[e.g.][]{elmegreen2010,guszejnov2018,ma2020}. In this scenario, UV-bright clumps are simply star cluster complexes formed in high-z galaxies, and thus likely trace the star-formation process in their host galaxy.

With respect to local stellar clusters and cluster complexes, high-redshift clumps are more extreme systems, usually with elevated star formation rate (SFR) and SFR densities \citep[e.g.][]{livermore2015,messa2022,claeyssens2023}, and sometimes observed as mini-starburst entities within their host galaxies \citep{zanella2015,iani2022}. Within the \textit{in-situ} formation scenario, this difference can be explained by high-z disks fragmenting at much larger scales (and possibly densities) than in local MS galaxies because of their gas-rich and turbulent nature, as suggested by models \citep[e.g.][]{immeli2004b}, numerical simulations of turbulent high-redshift galaxies \citep[e.g.][]{tamburello2015,renaud2021,vandonkelaar2022}, observations of dense giant molecular cloud complexes from CO data in galaxies at $\rm z=1$ \citep{dessauges2019,dessauges2023}, as well as by observations in nearby analogs \citep[e.g.][]{fisher2017a,fisher2017b,messa2019}.

While the \textit{in-situ} origin seems to explain the formation of $\sim70\%$ of the clumps at redshifts $\rm z<3$ \citep{zanella2019}, it is still possible that galaxies at earlier times were characterised by clumps formed \textit{ex-situ}, i.e. by a merger process during which the satellite is stripped and its nucleus appears as a massive clump \citep[e.g.][]{somerville2000,hopkins2008,puech2010,straughn2015,ribeiro2017}. This scenario could be justified by the increasing galaxy (minor and major) merger rate towards higher redshifts \citep[e.g.][]{lotz2011}, as well as by the increasingly lower number of massive galaxies \citep{marchesini2009}; simulations show that large-enough galaxy masses are needed for disk fragmentation to happen \citep[e.g.][]{tamburello2015}. 
Unfortunately systems before cosmic noon ($z\gtrsim4$), where violent disk instability is thought to have less impact on clump formation, are currently under-represented in clump studies \citep[e.g.][]{mestric2022}. The JWST will soon bring a new insight on clump formation at these redshifts, by providing deep, high-resolution rest-frame optical observations \citep[as seen from the first results by e.g.][]{mowla2022,claeyssens2022,vanzella2022_sunrise}. 

In this paper, we analyse the clump populations of three bright and highly magnified gravitationally-lensed galaxies at $\rm z>3.5$, in order to pave the way for upcoming JWST studies of larger samples. The selection, main properties and observational data of the sample are presented in Sect.~\ref{sec:data}; the clump analysis methodology is described in Sect.~\ref{sec:datanalysis}; results are presented in Sect.~\ref{sec:results} and discussed, also in the context of other literature works, in Sect.~\ref{sec:discussion}; finally, the main details of this analysis are summarised in Sect.~\ref{sec:conclusion}.
Throughout this paper, we adopt a flat $\rm \Lambda$-CDM cosmology with $H_0=68$ km s$^{-1}$ Mpc$^{-1}$ and $\rm \Omega_M = 0.31$ \citep{planck13_cosmo}, and the \citet{kroupa2001} initial mass function. All quoted magnitudes are on the AB system.

\section{Sample and Data}
\label{sec:data} 

\subsection{Galaxy Sample}
\label{sec:sample} 
\begin{figure}
    \centering
    \includegraphics[width=0.49\textwidth]{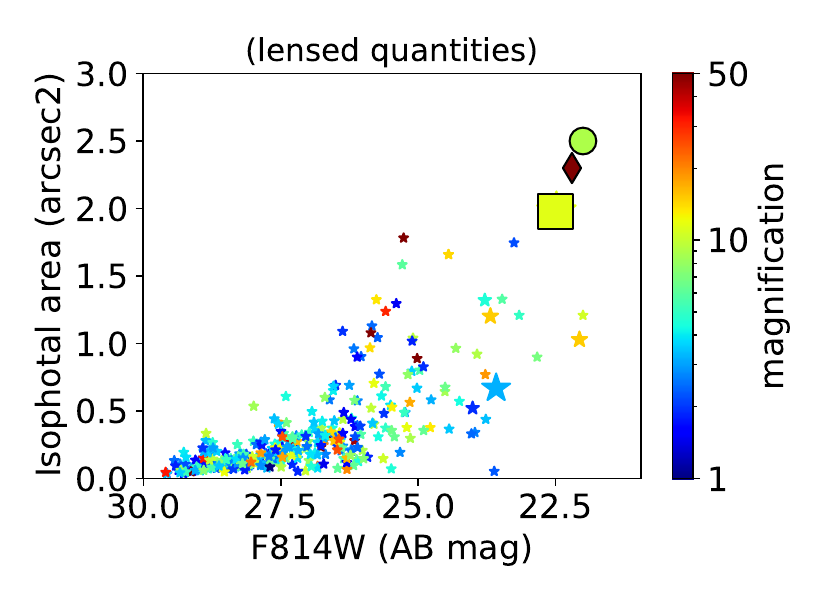}
    \caption{Observed (lensed)} rest-frame UV magnitudes and angular area covered by spectroscopically confirmed $z>3.5$ galaxies gravitationally lensed by galaxy clusters taken from the literature (see text for details). The sizes of the markers is proportional to the number of clumps observed in the galaxy. The final selection of 3 galaxies analysed in the current work are highlighted by the following markers: a circle (MS1358), a diamond (RCS0224) and a square (MACS0940).    
    \label{fig:sample_sel}
\end{figure}

We search for spatially extended lensed galaxies at $z>3.5$, with spectroscopically-confirmed redshifts. These criteria are chosen to characterise sub-galactic scales at redshifts where, as outlined in the introduction, we may expect different formation mechanisms than cosmic noon.
The search is restricted to cluster lensing, to allow for larger magnifications across the full extent of the arcs. The sample of clusters is taken from the full MAssive Clusters Survey (MACS) survey (e.g. \citet{ebeling2010,repp2016}), as well as other clusters taken from the LoCuSS \citep{richard2010}, CLASH \citep{postman2012}, Frontier Fields \citep{lotz2017} and RELICS \citep{coe2019} cluster samples. 
In Fig.~\ref{fig:sample_sel} we observe how, while the typical population of background lensed galaxies behind many massive galaxy clusters is distributed around apparent (and magnified) rest-frame UV magnitudes in the range $26-30$ mag (see also \citealp{richard2021}, \citealp{claeyssens2022}), three galaxies clearly stand out in terms of brightness and angular area.
Those galaxies are located beyond the lensing clusters MS 1358+62 (at $\rm z_{cl}=0.33$), RCS 0224-0002 (at $\rm z_{cl}=0.77$) and MACS J0940.9+0744 (at $\rm z_{cl}=0.34$); the lensed systems are in the redshift range $\rm z=4-5$ and are all magnified by large factors ($\mu>5$).
The selected targets are typical of their redshift, in terms of masses and SFRs (see Sections \ref{sec:MS1358}, \ref{sec:RCS0224} and \ref{sec:MACS0940})
but appear ‘clumpy’, i.e. host more than one bright clump (in Fig.~\ref{fig:sample_sel}, the marker size is proportional to the number of observed clumps), due to their large magnification factors, making them optimal candidates for the study of star formation at small sub-galactic scales, down to 10 pc (see Section~\ref{sec:results_phot}).
In addition, their large magnified observed brightness makes them ideal targets for future JWST observations\footnote{JWST observations with NIRSpec-IFU and NIRCam imaging have been approved for these targets, GO program 3433.}.

We summarise below the main properties of these galaxies, from previous studies in the literature; for simplicity, we will use the shortened names of the galaxy clusters as the names of the high-z lensed galaxies we analyse in this work.

\subsubsection{MS1358}\label{sec:MS1358}
The $z=4.92$ lensed galaxy behind MS 1358+62 galaxy cluster was first discovered and studied by \citet{franx1997}. The galaxy rest-frame UV/optical morphology is dominated by several compact star-forming regions. Fitting the broad-band SED from HST and \textit{Spitzer} imaging, \citet{swinbank2009} derived a total mass $\rm M_\star=\left(4.7\pm1.3\right)\times10^8\ M_\odot$\footnote{\label{note1}The original values of mass ($\rm M_\star=7\times10^8\ M_\odot$) and star formation rate ($\rm SFR=42\ M_\odot yr^{-1}$ for the entire galaxy and $\rm SFR=18\ M_\odot yr^{-1}$ for the two brightest clumps) were derived by \citet{swinbank2009} assuming a \citet{salpeter1955} IMF and are here converted to match the assumption of \citet{kroupa2001} IMF used throughout the current paper.}, from a young stellar population ($\rm 14\pm7\ Myr$) with sub-solar metallicity ($\rm Z=0.2\ Z_\odot$); the stellar extinction is consistent with being close to zero, $\rm E(B-V)=0.05\pm0.05$ mag. A star formation rate of $\rm SFR=28\pm5\ M_\odot yr^{-1}$ was derived from the [O\protect\scaleto{$II$}{1.2ex}] emission-line flux \citep{swinbank2009}. 
The galaxy is characterised by the presence of two main sub-galactic star-forming regions (IDs: 1 and 2 in Fig.~\ref{fig:galaxies1}) accounting for $\rm 12\pm1\ M_\odot yr^{-1}$ in SFR. 
IRAM PdBI observations in CO(5–4) emission suggest that the galaxy has a large gas fraction, $\rm f_{gas}=0.6$ \citep{livermore2012b}, similarly to what is observed in $\rm z\geq3$ galaxies \citep{dessauges2020}. 

\subsubsection{RCS0224}\label{sec:RCS0224}
A lensing-magnified arc at $z=4.88$ in the RCS 0224-0002 cluster, it was first discovered as a bright \lya\ halo by \citet{gladders2002}. \citet{swinbank2007} studied the rest-frame ultraviolet and optical properties of the galaxy by combining HST imaging to VIMOS and SINFONI spectroscopy, deriving a dynamical mass of $\rm \sim10^{10}\ M_\odot$ (within the estimated 2 kpc intrinsic size of the galaxy) and a star formation rate of $\rm 8.2\pm1.4\ M_\odot yr^{-1}$ (using [O\protect\scaleto{$II$}{1.2ex}] observations\footnote{This value has been adapted from the original $\rm 12\pm2\ M_\odot yr^{-1}$ to translate from \citet{kennicutt1998araa} to the \citet{kennicutt2012} calibrations.} covering only the central and western, C and W, images in Fig.~\ref{fig:galaxies2}). A further study with MUSE suggested that the detected emission lines are powered by a young ($<5$ Myr) and low-metallicity ($\rm Z\lesssim0.05\ Z_\odot$) stellar population \citep{smit2017}.

\subsubsection{MACS0940}\label{sec:MACS0940}
MACS0940 at $\rm z=4.03$, first studied by \citet{leethochawalit2016}, is observed as a strongly stretched arc, made of two mirrored images (see Fig.~\ref{fig:galaxies3}). Two complete counter-images of the same galaxy are seen northern and eastern from the arc. The galaxy is characterised by a bright Ly$\alpha$ halo \citep{claeyssens2019}, suggesting recent episodes of star formation. Bright compact sources are clearly visible along the arc in rest-frame UV observations. 

\subsection{Hubble Space Telescope (HST) observations}
\label{sec:data_hst}
HST observations with WFPC2, ACS/WFC and WFC3/IR are available on the HST MAST archive (see Data Availability section at the end of the publication for details about program IDs). The list of filters used is given in Tab.~\ref{tab:galaxies}. For each galaxy, we choose a reference filter, covering the rest-frame UV wavelengths (see Tab.~\ref{tab:ref_filters}), where we measure the size and UV luminosity of the clumps (as described in Section~\ref{sec:modeling}); the other filters are used to infer the broad-band SED of the clumps (see Section~\ref{sec:BBSED}). 
Individual flat-fielded and CTE-corrected exposures were aligned and combined in single images using the \texttt{AstroDrizzle} procedure from the \texttt{DrizzlePac} package \citep{hoffmann2021}. 
The astrometry was aligned to the Gaia DR2 \citep{gaia2018}.

We model the instrumental point-spread function (PSF) from isolated bright stars within the field of view of the observations. In each filter, we fit the selected stars with an analytical function described by a combination of Moffat, Gaussian, and $\rm 4^{th}$ degree power-law profiles. Such a generic combination is chosen to mitigate possible bias introduced by the choice of a specific function. The fits provide good descriptions of the stars up to a radius of $\sim0.8''$, i.e. more than $\rm 10\times$ larger than the half-width at half maximum (HWHM) of PSF in the reference filters (the full width at half maximum, FWHM, of the reference filters are $0.10''$, $0.11''$ and $0.12''$ for MS1358, RCS0224 and MACS0940, respectively, see Tab.~\ref{tab:ref_filters}). The modeled PSF is used to derive the size and luminosity of the clumps, as described in detail in Section~\ref{sec:modeling}.
\begin{table*}
    \centering
    \begin{tabular}{lcrrlll}
        \multicolumn{1}{c}{Galaxy} & \multicolumn{1}{c}{z} & \multicolumn{1}{c}{$\rm N_{img}$} & \multicolumn{1}{c}{$\rm \mu_{range}$}  & \multicolumn{1}{c}{Lens model ref.} & \multicolumn{1}{c}{$\rm F_{ref}^\dag$} & \multicolumn{1}{c}{Other filters$^\dag$} \\
        \multicolumn{1}{c}{(1)} & \multicolumn{1}{c}{(2)} & \multicolumn{1}{c}{(3)} & \multicolumn{1}{c}{(4)} & \multicolumn{1}{c}{(5)} & \multicolumn{1}{c}{(6)}  & \multicolumn{1}{c}{(7)} \\
    \hline
    \hline
        MS1358   & 4.92 & 3 (1) & \ 3-20 & \citet{swinbank2009} & F775W$^\ddag$ & F625W, F814W, F850LP, F110W, F160W \\
        RCS0224  & 4.88 & 3 (1) & 30-85 & \citet{swinbank2007}, this work & F814W & F606W, F125W, F160W \\
        MACS0940 & 4.03 & 4 (0) & \ 6-33 & \citet{claeyssens2019, richard2021}& F814W & F606W, F125W, F160W  \\
    \hline
    \end{tabular}
    \caption{Galaxy sample and HST data: (1) ID of the galaxy (we remind that we are using the name of the hosting galaxy clusters as names/IDs of the lensed arcs studied); (2) redshift of the galaxies, from \citet{franx1997}, \citet{gladders2002} and \citet{leethochawalit2016}; (3) number of lensed images analysed (in parenthesis the number of counter-images not included in the clump study); (4) range of magnification values covered by the galaxy; (5) references for the lens models; (6) reference filter used to extract the size and UV magnitude of the clumps; (7) other filters, used to derive the broadband SED of the clumps. 
    $^\dag$All filters are from either WFC3 or ACS, except F606W from WFPC2. 
    $^\ddag$F775W is used instead of F814W as the reference filter for MS1358 because of its longer exposure time, providing better signal-to-noise.
    }
    \label{tab:galaxies}
\end{table*}

\begin{table}
    \centering
    \begin{tabular}{lcccccc}
        \multicolumn{1}{c}{Galaxy} &  $\rm F_{ref}$ & $\rm \lambda_{rest}$ & $\rm t_{exp}$ & $\rm PSF$ & $\rm lim_{ext}$ & $\rm lim_{com}$ \\ 
        \ & \ & [\AA] & [s] & [arcsec] & [mag] & [mag] \\ 
        \multicolumn{1}{c}{(1)} & \multicolumn{1}{c}{(2)} & \multicolumn{1}{c}{(3)} & \multicolumn{1}{c}{(4)} & \multicolumn{1}{c}{(5)} & \multicolumn{1}{c}{(6)}  & \multicolumn{1}{c}{(7)} \\
        \hline
        \hline
        MS1358 & F775W & 1300 & 5470 & 0.10 & 27.7 & 27.9 \\
        RCS0224 & F814W & 1370 & 2168 & 0.11 & 26.5 & 27.1 \\ 
        MACS0940 & F814W & 1600 & 7526 & 0.12 & 27.3 & 27.4 \\ 
        \hline
    \end{tabular}
    \caption{Main properties of the observations in the reference filter for each galaxy: (1) galaxy ID; (2) reference filter; (3) rest-frame pivotal wavelength; (4) exposure time; (5) \& (6) extraction and completeness limits for a point-like source within the galaxy, as described in Section~\ref{sec:completeness}; note that these values are corrected for the Galactic reddening.}
    \label{tab:ref_filters}
\end{table}
%
%
\subsection{Gravitational lens models}
\label{sec:lens_model}

Our study relies on the reconstruction of intrinsic properties of the identified clumps, accounting for the anisotropic lensing magnification which is estimated from a lens model.

As a starting point, we make use of existing mass models of the cluster cores (references in Table \ref{tab:galaxies}) which include the lensed arcs as constraints. In summary, these models use a parametric mass distribution describing the cluster-scale and galaxy-scale mass components of the clusters as a combination of double Pseudo Isothermal Elliptical (dPIE) profiles. The models are optimised with the Lenstool \citet{jullo2007} software\footnote{publicly available at \url{https://projets.lam.fr/projects/lenstool/wiki}} based on multiple imaged constraints, similarly to the modelling performed in e.g. \citet{Richard2014}. In the case of the older models for MS1358 and RCS0224, we have performed a new optimisation using the latest version (v8) of Lenstool. 

In all cases, individual clumps identified as mirrored images in the three lensed arcs are individually used as constraints in the model, ensuring a very good reproduction of the images. The rms between the observed and predicted locations of the images used in the model is 0.06", 0.08" and 0.24" for MS1358, RCS0224 and MACS0940 respectively.

The outcome of the Lenstool optimisation is a statistical sample of mass models sampling the posterior distribution function of the model parameters. We make use of both the best model (achieving the lowest rms) and the range of mass models to estimate the magnification factors  (along each direction) at each position across the arc and their 68\% percentile error. The range of magnification factors obtained is summarised in Table \ref{tab:galaxies}.

\begin{figure*}
    \centering
    \includegraphics[width=\textwidth]{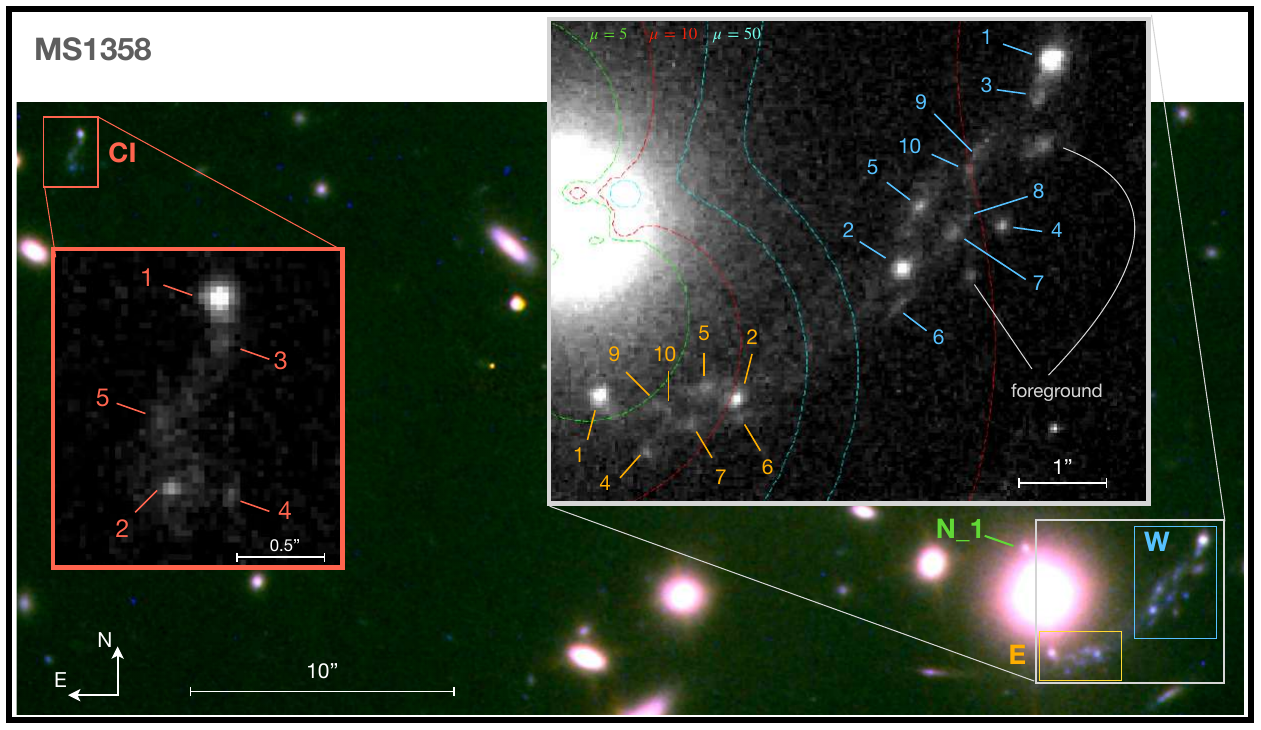}
    \caption{RGB composite (r:F160W, g:F110W, b:F775W) of the galaxy cluster field MS 1358+62, containing the $z=4.93$ lensed western (W) and eastern (E) arcs and their counter-image (CI). The zoomed-in inset, showing the observations in the reference F775W filter, highlights the positions and names of the extracted stellar clumps in all the different multiple images. A partial northern image (N) of the lensed galaxy contains only one clump and is labeled as `N\_1'.
    Dashed lines show regions of constant magnifications at $\mu=5$, 10 and 50 (green, red and cyan, respectively), at the redshift of the lensed arcs. Reference angular scale are given; the image is aligned with north-up and east-left.}
    \label{fig:galaxies1}
\end{figure*}
\begin{figure*}
    \centering
    \includegraphics[width=\textwidth]{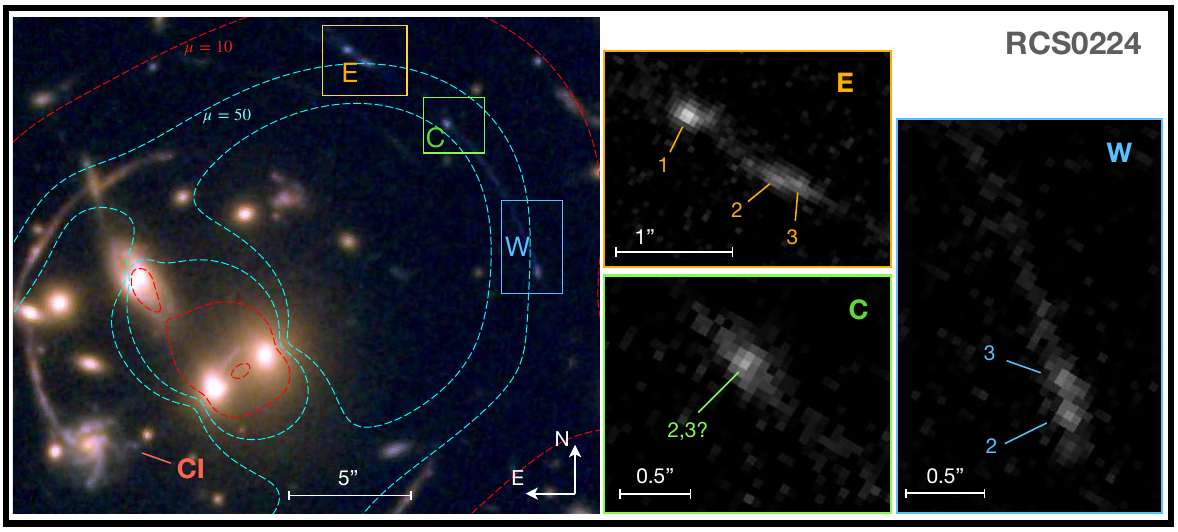}
    \caption{Same as Fig.\ref{fig:galaxies1}, for RCS0224 (r:F160W, g:F125W, b:F814W, zoom-ins:F814W). According to the lens model the clump seen in the central (C) image is consistent with the position of either clump 2 or 3; the study of clump properties suggests that what is observed is a counter-image of clump 2 (see Section~\ref{sec:discussion_individual} and Tab.~\ref{tab:results}).}
    \label{fig:galaxies2}
\end{figure*}
\begin{figure*}
    \centering
    \includegraphics[width=\textwidth]{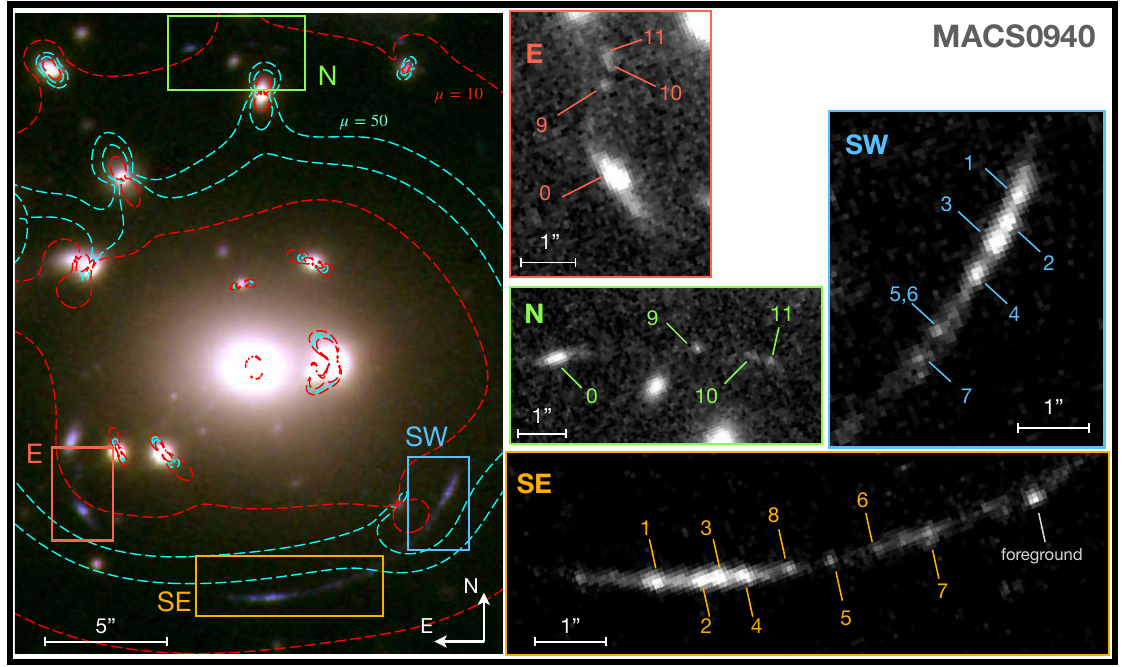}
    \caption{Same as Fig.\ref{fig:galaxies1}, for MACS0940 (r:F160W, g:F125W, b:F814W, zoom-ins:F814W). According to the lens model of the system, the SE and SW arcs (and all the clumps they host) are different lensed images of the main galaxy (labeled as ``0'' in the N and E fields). The clumps 9, 10 and 11, observed in the N and E fields, do not have counterparts in the SE and SW fields.}
    \label{fig:galaxies3}
\end{figure*}
\section{Data analysis}
\label{sec:datanalysis}

\subsection{Clump extraction}
\label{sec:sextraction}
Clump candidates have been extracted by running \texttt{SExtractor} \citep{bertin1996} requiring $3\sigma$ detections in at least 4 pixels (px) in background subtracted images (where the background is evaluated on a 30 px scale) in the reference filter. In order to check for `redder' sources that could be missed in the reference filter, we ran \texttt{SExtractor} with the same set of parameters also in all the other available filters in each galaxy; this test did not produce any further source. 
The colors of the clumps and the lens models have been used to identify interlopers, mainly bright compact sources at different redshifts, among the extracted sources (some examples are highlighted in Figs.~\ref{fig:galaxies1} and~\ref{fig:galaxies3}). 

The extracted clumps are shown, together with their assigned IDs, in Figs.~\ref{fig:galaxies1}, \ref{fig:galaxies2} and \ref{fig:galaxies3}. The lens models predictions were used to match the same clump over different images. Ten and 8 clumps are observed in the western (W) and eastern (E) arcs of MS1358, respectively; only 5 of them are clearly visible in the counter-image (CI). The brightest clump (ID:1) is the only one visible in the northern image (N\_1).
RCS0224 appears in the eastern image (E) as composed of two bright regions, one of which is composed by two sub-clumps. Only one of such regions (containing clumps 2 and 3), is visible in the western image (W); the central image (C) is instead characterised by a single source (either clump 2 or 3), with a very high magnification. The counter-image (CI) of RCS0224 appears as a single source, where clumps are indistinguishable.
MACS0940 counts a total of 8 clumps in the eastern arc (SE), the one with the largest magnification overall. Only 6 of those clumps are seen in the western arc\footnote{Due to the uncertainties in the clumps predicted positions given by the lens model, one source is consistent with being the counterpart of either SE\_5 or SE\_6 and was labelled SW\_5,6 to reflect this.} (SW), while both the eastern and northern counter-images (E and N) look like a single bright source (consistent with the superposition of the brightest clumps, IDs: 1, 2, 3 and 4) surrounded by diffuse light and by 3 sources (IDs: 9, 10 and 11, not seen in the arcs).

\subsection{Clump modeling}
\label{sec:modeling}
Clumps are modelled on the image plane, fitting their sizes and fluxes on the observed data of the reference filter; such quantities are later converted into intrinsic values, using the magnification maps produced by the lens model (Section~\ref{sec:lens_model}).
The clump modelling follows accurately the methodology applied to the study of stellar clumps in the z=1 galaxy A521-sys1 (Section 3.2 of \citealp{messa2022}) of which we summarize here the main features. We assume that clumps have Gaussian profiles in the image plane, 
and therefore we fit a model consisting of 2D Gaussian functions, convolved to the the instrumental point spread function (i.e. the response of the instrument), to obtain their observed profiles. 

We performed clump fitting in $9\times9$ px cutouts centered on each of the clumps, where a 1st degree polynomial function is added to account for the galaxy background luminosity: 
both size (including axis-ratio and orientation, for elliptical sources) and flux of the clumps are left as free parameters of the fit in the reference filter. 
The best-fit models and residuals are shown in Appendix~\ref{sec:app:plots_bestfit}.
The clump shape is then kept fixed in the other filters, where only the source flux is fitted, under the assumption that the intrinsic clump shape is the same in all bands. 
In order to reduce the possible contamination from nearby bright sources, the brightest clumps were fitted first, and then their best-fit model subtracted from the data.
The fitted fluxes, in units of $e^-/s$, are converted into observed AB magnitudes by considering the instrumental zeropoints and by subtracting the reddening introduced by the Milky Way in each of the filters.
The (observed) effective radius of the clumps, $\rm R_{eff,obs}$, here defined as the radius enclosing half of the source's luminosity, is equal to the (circularized) full width at half maximum, in the current assumption of gaussian profiles.
In the occurrence of multiple clumps separated by less than 5 px, a single fit is performed in cutouts large enough to include all sources (typically $15\times15$ px).

As already discussed in \citet{messa2019} and \citet{messa2022}, the oversampling of the HST PSF (FWHM$\sim2$ px in the filters considered in this work) allow to resolve sub-pixel clump sizes. By inserting mock clumps of various sizes in the image frames and fitting them as for the real clumps, we derive the minimum resolvable size, $\rm \sigma_{min} = 0.40$ px for MS1358 and MACS0940, and $\rm \sigma_{min} =0.45$ px for RCS0224. These limits, converted into physical sizes at the redshift of the galaxies, are highlighted in the top panels of Fig.~\ref{fig:phot_results}.
We refer to Appendix C of \citealp{messa2022} for more details on the process of estimating the minimum resolvable size.

\subsubsection{Completeness of the sample}
\label{sec:completeness}
We separately tested the completeness we reach in extracting and analysing our samples. 
In order to test the sensitivity of our extraction process we insert, one by one, 1500 synthetic clumps at random positions within the region covered by the lensed systems (in the reference filter used in the original extraction, Section~\ref{sec:sextraction}). The clumps are modelled as symmetric gaussians and divided into 3 `size groups', with $\sigma=0.4$ ($0.45$ for RCS0224), $1.0$ and $2.0$ px; the first value correspond to the minimum resolvable size (see Section~\ref{sec:modeling}), while the last encompass the largest clumps observed in the samples (see Fig.~\ref{fig:phot_results}, top row). 
Clumps are randomly drawn from a flat distribution in magnitudes. 
For each realisation we run \textsc{SExtractor} with the same sets of parameters used in Section~\ref{sec:sextraction} and we check if the synthetic clump has been extracted. For each of the three sizes chosen we consider as extraction limit, $\rm lim_{ext}$ the magnitude below which the fraction of extracted clumps goes above $90\%$.

In order to test the reliability of the derived photometry, we estimate the completeness of the sample in a second way, following the method described in Appendix~D of \citet{messa2022}; the same synthetic clumps used to estimate the extraction limits are photometrically analyzed in the same way as the real sources. We consider as good sources the ones where the relative error on the recovered flux is below $50\%$, $\rm flux_{rel} = |flux_{in}-flux_{out}|/flux_{in}<0.5$. We consider as completeness limits, $\rm lim_{com}$, the magnitudes below which the fraction of good sources recovered goes above $80\%$.
All the extraction and completeness limits are compared to the photometry of the clumps in the sample in Fig.~\ref{fig:phot_results} (top row). The limits in the case of point-like clumps are also reported in Tab~\ref{tab:ref_filters}.

\subsection{Conversion to intrinsic sizes and magnitudes}\label{sec:convert_intrinsic}
In order to recover the intrinsic clumps' properties, we considered the magnification map produced by the best fit lens model of each galaxy; we used as reference amplification, $\rm \mu_{tot}$, the median amplification value found in the region centered on each clump coordinates and extending within one FWHM of its size. We use the standard deviation of amplification values found in the same region as an estimate of the magnification uncertainty associated to the clump position.
To estimate the uncertainty associated to the magnification map, we consider, for each cluster, the $1\sigma$ interval ($\rm 16^{th}$ to $\rm 84^{th}$ percentiles) of the amplification values found in the 500 lens model variations described in Section~\ref{sec:lens_model}. The two magnification uncertainties are combined (by the sum of the squares) into a final uncertainty. We note that the uncertainty from the 500 `variation' maps is the one usually dominating the final value. 

The observed magnitudes are converted into absolute ones by subtracting the distance modulus and adding the $k$ correction, a factor $\rm 2.5\log(1+z)$; the amplification is accounted for by adding $\rm 2.5\log(\mu_{tot})$, converting the magnitude into the intrinsic value. 

We consider three cases when measuring the intrinsic effective radius, $\rm R_{eff}$, following the methodology already used and discussed in the literature \citep[e.g.][]{vanzella2017a,claeyssens2023}. If the clump is resolved along both minor and major axis in the image plan, we simply divide the observed $\rm R_{eff,obs}$ by the square-root of the clump amplification. If the clump is un-resolved along the transversal direction of the magnification, we consider as informative only the size measured in the shear direction, and we divide the latter by the tangential component of the magnification ($\rm \mu_{tan}$). Finally, if the clump is un-resolved in both directions we divide the observed $\rm R_{eff,obs}$ (consistent with the instrumental PSF) by $\rm \mu_{tan}$ to derive a size upper limit. In the last two cases, the underlying assumption is that the clump has approximately a circular shape in the source plane. 
The final uncertainties on the intrinsic properties combine both photometric and magnification uncertainties via the root sum squared.

\subsection{Broadband SED fitting}
\label{sec:BBSED}
We use the broadband photometry to estimate the clumps' masses. Following the methodology of \citet{messa2022} and \citet{claeyssens2023},
photometry in the filters other than the reference one is performed by assuming that each source has the same observed size (derived in Section~\ref{sec:modeling}) in all the bands, and thus fitting only the flux and the background.
Due to the elevated redshift of the sources, the available HST filters cover only the rest-frame wavelength range $\sim1000-3000$ \AA. As a consequence our SED-derived values are UV-weighted quantities. In particular, ages and extinctions are only poorly constrained; on the other hand, mass estimated are more robust, and within a factor of few correct (see also \citealp{messa2022} for the robustness of mass estimates over different model assumptions).

To mitigate the effects of degeneracies between parameters of the fit (in particular ages, extinctions, metallicities and star formation histories, SFHs), we limit the number of free parameters. 
First of all, we fixed the metallicity of the stellar models; following \citet{stark2009} we use a sub-solar metallicity, $\rm Z=0.2\ Z_\odot$. \citet{smit2017} suggest that the nebular \CIV\ lines observed in RCS0224 must come from a young stellar population with metallicity $\rm Z=0.05\ Z_\odot$ or lower. For this reason we perform, for all galaxies, a second `control' fit using stellar models with metallicity $\rm Z=0.02\ Z_\odot$.
As second strong assumption, we consider SFHs described by a 10 Myr continuous star formation. This choice is driven by the sizes of the clumps (Section~\ref{sec:results_phot}), larger than typical stellar clusters (for which an instantaneous burst is usually assumed); assuming longer histories, like a 100 Myr continuous star formation, would lead to larger masses on average (as already found by, e.g. \citealp{adamo2013,messa2022,claeyssens2023}).
We use the stellar models from the \texttt{Yggdrasil} stellar population synthesis code\footnote{Yggdrasil models can be found at \url{https://www.astro.uu.se/~ez/yggdrasil/yggdrasil.html}} \citep{zackrisson2011}, based on Starburst99 Padova-AGB tracks \citep{leitherer1999,vazquez2005} with a universal \citet{kroupa2001} IMF (in the mass interval $\rm 0.1-100\ M_\odot$), processed through \texttt{Cloudy} \citep{ferland2013}, assuming a $50\%$ nebular covering fraction, to obtain the evolution of the nebular continuum and line emission produced by the ionized gas. 
The model spectra, at each age, are attenuated with a color excess ranging between $\rm E(B-V)=0-1$ mag, using the \citet{calzetti2000} law, before being convoluted with the filter throughput. We do not use the differential expression formulated by \citet{calzetti2000}, but we apply the same reddening to both stellar and nebular emission, assuming that stars and gas are well mixed. We check \textit{a-posteriori} that the accepted solutions for the SED fit have low extinctions, $\rm E(B-V)\leqslant0.4$ mag in most cases.
Age, mass and extinction are left as free-parameters in the SED fitting process. Best-fit parameters are given by the model with the lowest reduced $\rm \chi^2$ ($\rm \chi^2_{red.,min}$). Their uncertainties are given by the entire range of solutions whose reduced $\rm \chi^2$ satisfies $\rm \chi^2_{red.}\leq 2\times\chi^2_{red.,min}$; this value was chosen, by inspecting the fit results, as it encompasses similarly-good solutions.
Like it was the case for magnitudes in Section~\ref{sec:convert_intrinsic}, de-lensed masses are derived by dividing the observed mass of each clump by its total magnification.

\section{Results}
\label{sec:results}
\subsection{Clumps sizes and luminosities}
\label{sec:results_phot}
\begin{figure*}
    \includegraphics[width=0.33\textwidth]{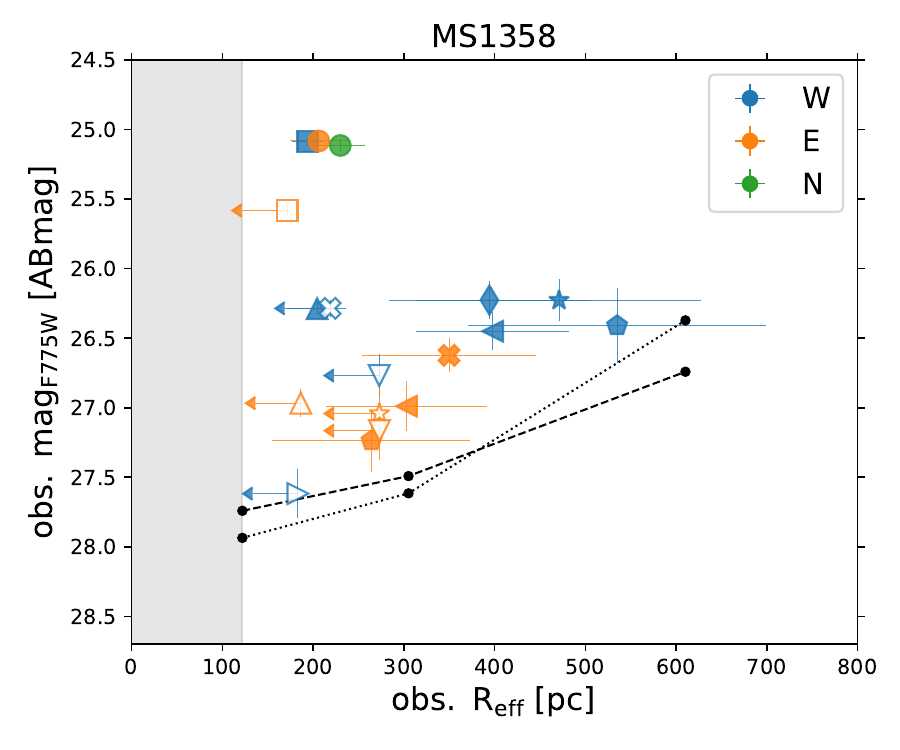}
    \includegraphics[width=0.33\textwidth]{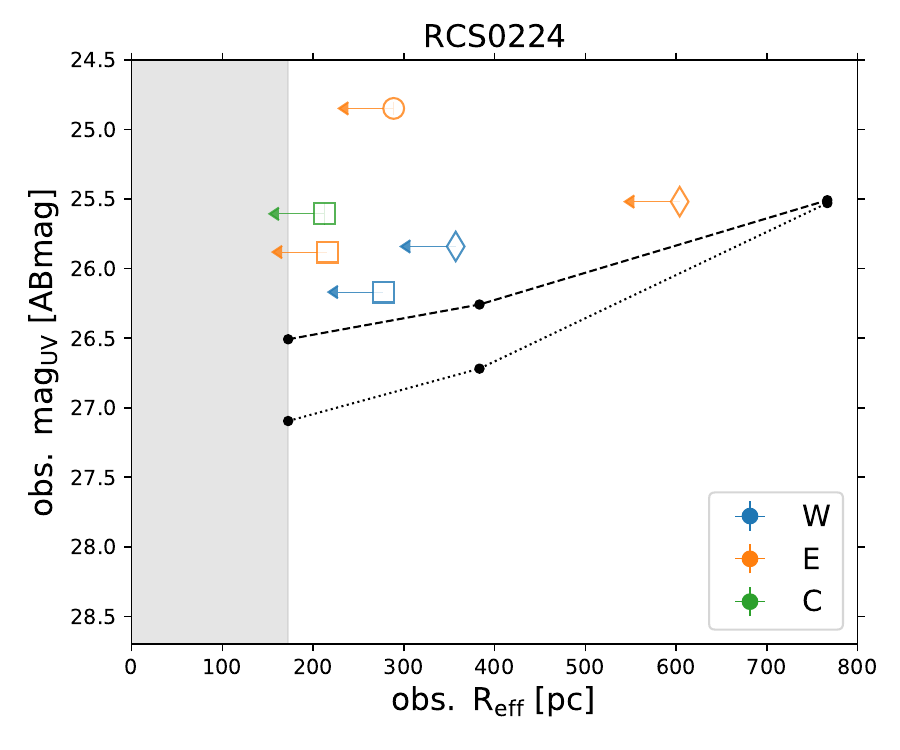}
    \includegraphics[width=0.33\textwidth]{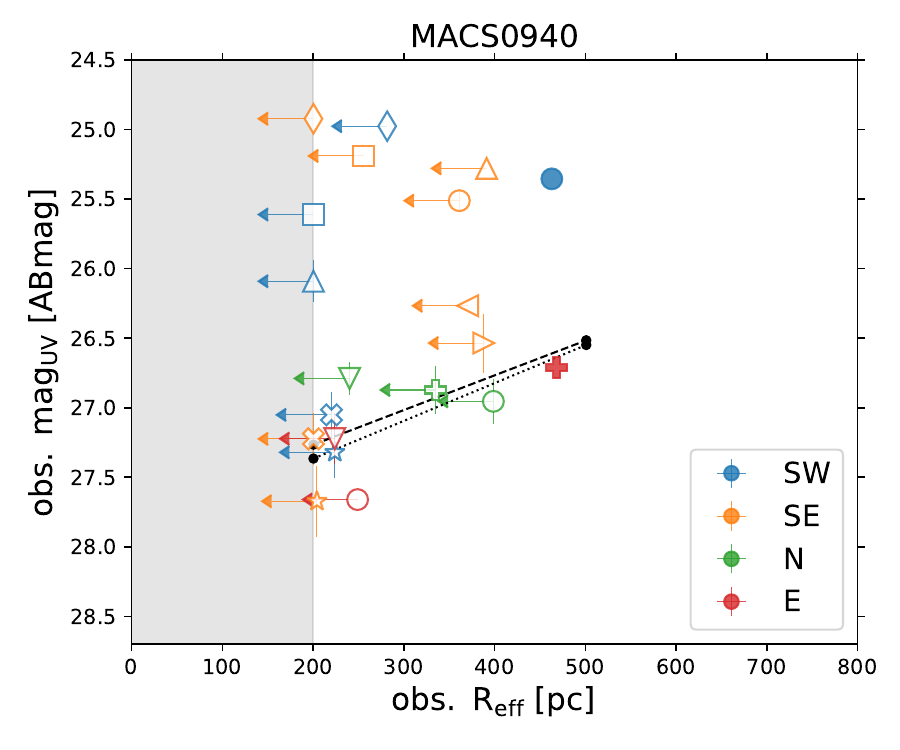}
    \includegraphics[width=0.33\textwidth]{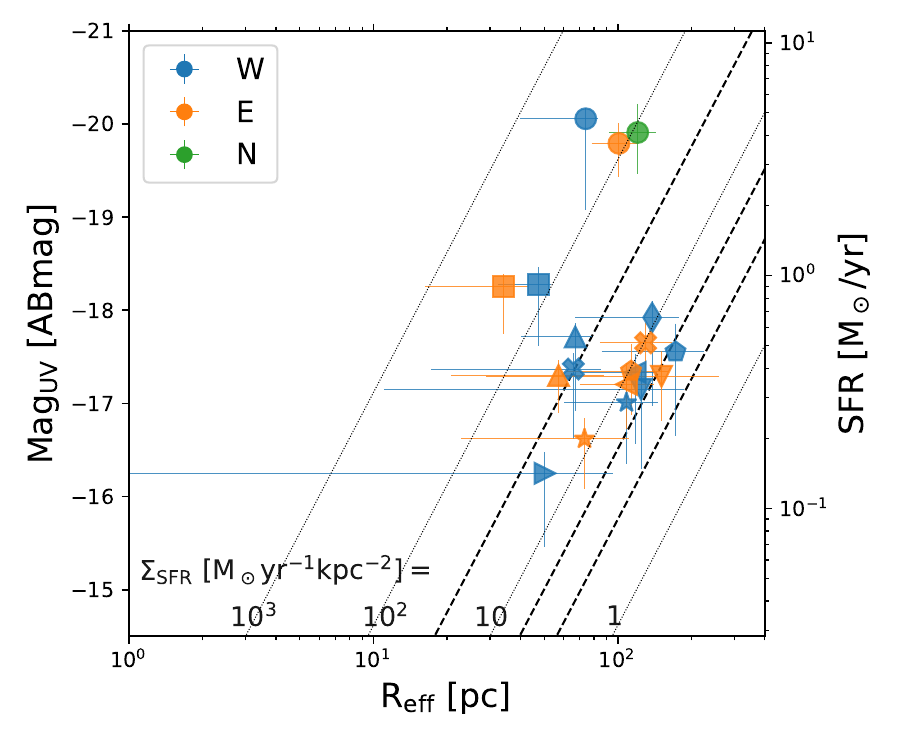} 
    \includegraphics[width=0.33\textwidth]{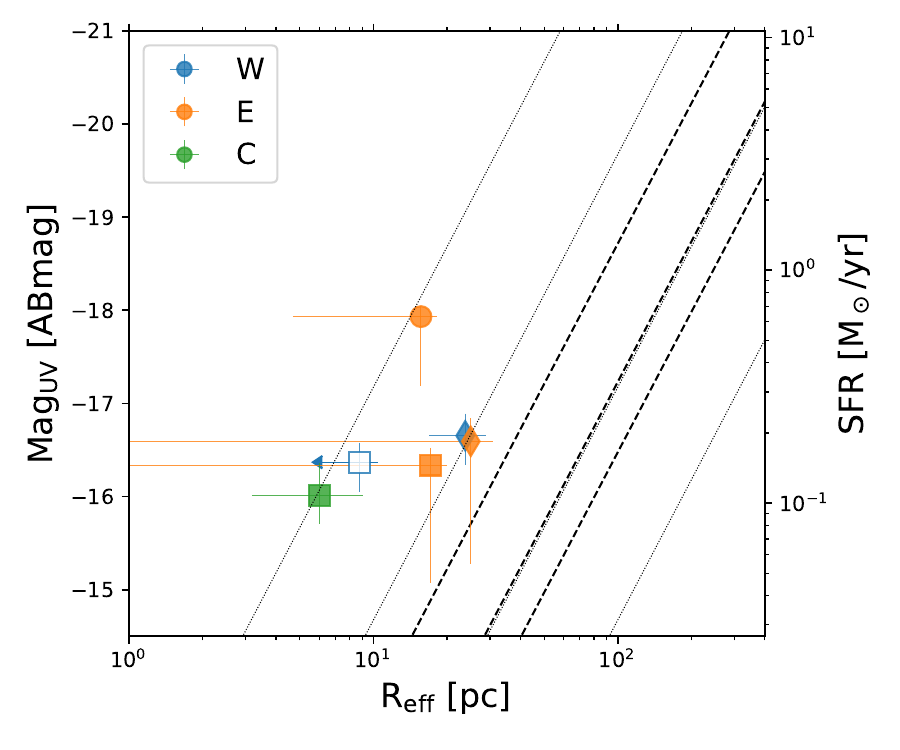}
    \includegraphics[width=0.33\textwidth]{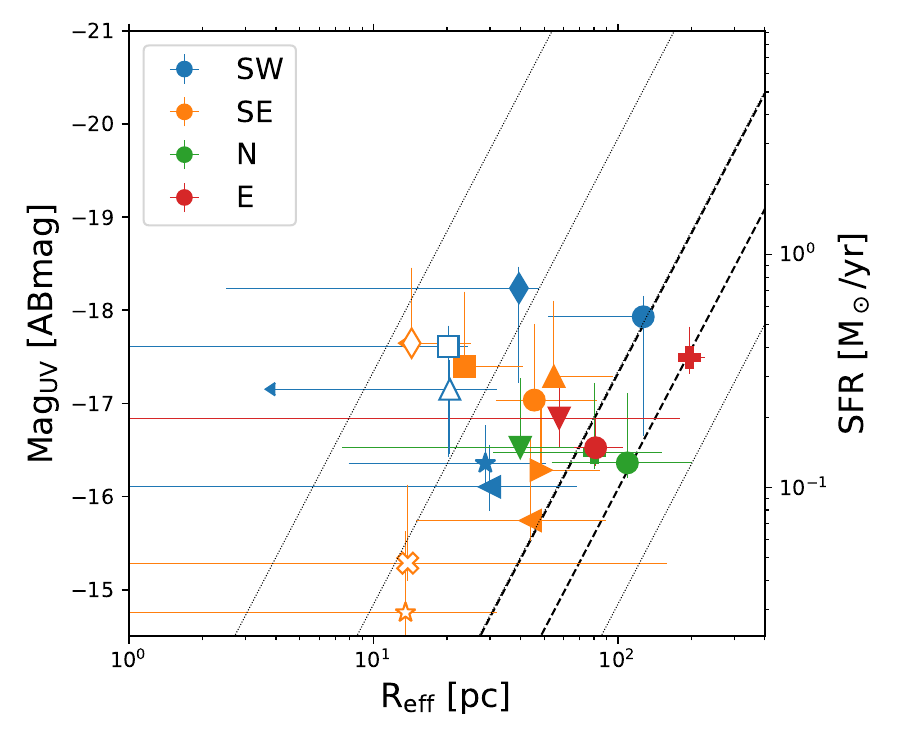}
    \caption{\textit{(Top panels:)} observed sizes and magnitudes (in the reference filter) for all the clumps in this study. Estimates of the observed sizes consider the angular diameter distance of the host galaxy but do not take into account their lensing magnification. The gray shaded areas highlight the size regions below our resolution limits. Black dot markers (connected by dashed lines) indicate the extraction limits for each of the simulated sizes considered (see main text in Section~\ref{sec:completeness} for details); similarly, markers connected by dotted lines are used to indicate completeness limits.
    \textit{(Bottom panels:)} Intrinsic sizes and magnitudes of the clumps. UV magnitudes have been converted into $\rm SFR_{UV}$ values using the \citet{kennicutt2012} relation on the right-hand side of each panel. Uncertainty bars combine both photometric and lensing uncertainties. The extraction limits shown in the top panels have been converted here to surface brightness limits (dashed lines, each line corresponds to the limit derived for one simulated size). Solid gray lines are of constant SFR surface density for $1$, $10$, $10^2$ and $10^3$ $\rm M_\odot yr^{-1} kpc^{-2}$. In all panels, open symbols are used for upper limits and same markers are used for the same clumps in different counter-images.}
    \label{fig:phot_results}
\end{figure*}

The observed (i.e. not de-lensed, and therefore not `intrinsic') clump sizes and magnitudes are shown in the top row of Fig.~\ref{fig:phot_results}, where they are compared to the extraction and completeness limits described in Section~\ref{sec:completeness}. The magnitude ranges are similar among the different galaxies ($\sim28-25$ AB mag) and, in the great majority of cases, the analyzed clumps are above the completeness limit, suggesting that their photometry is robust. The shallower extraction and completeness limits in RCS0224 (consistent with the exposure time of this galaxy being the shortest, see Tab~\ref{tab:ref_filters}) causes the average clump magnitude in this galaxy to be bright ($\sim 26-25$ mag). This result may suggest that we are missing  clumps at lower surface brightness.
A shaded-gray area in the plot marks the region below the size lower limits discussed in Section~\ref{sec:modeling}; in the absence of gravitational lensing we wouldn't be able to study clumps below $\sim100-200$ pc scales. 

The intrinsic (i.e. de-lensed) clump sizes and magnitudes are shown in the bottom row of Fig.~\ref{fig:phot_results} and are reported in Tab.~\ref{tab:results}.
Intrinsic sizes are smaller than $200$ pc and in many cases reach values close to 10 pc; in particular, due to the magnification factors associated to our galaxies, we are studying clumps at scales comparable to large individual star clusters in RCS0224 ($\rm R_{eff}=5-25$ pc) and slightly larger scales in the other two cases, consistent with the sizes of compact star-forming regions ($\rm R_{eff}=13-200$ pc in MACS0940 and $\rm R_{eff}=30-150$ pc in MS1358).
The rest-frame UV magnitudes, on the y-axis of Fig.~\ref{fig:phot_results} (bottom panels), have also been converted into a SFR value (using the conversion factors from \citealp{kennicutt2012}), as this is a commonly used parameter in high-z clump studies. 
In the same panels we show lines of uniform SFR surface densities ($\rm \Sigma_{SFR} = 1, 10, 10^2, 10^3\ M_\odot yr^{-1} kpc^{-2}$). The extraction limits described above, when translated into surface-brightness limits (dashed lines), correspond to $\rm \Sigma_{SFR}\sim10\ M_\odot yr^{-1} kpc^{-2}$. In all cases the completeness is above the typical $\rm \Sigma_{SFR}$ values of clumps in local main sequence galaxies $\rm \sim0.6\ M_\odot yr^{-1} kpc^{-2}$ \citep{kennicutt2003,livermore2015}. \\
In some cases, we are not able to robustly constrain the clump properties, mainly due to very large uncertainties on the magnification. 
Those cases are discussed for each galaxy individually in Section~\ref{sec:discussion_individual} and are then removed from following analyses.
A detailed discussion of clump intrinsic properties for each of the galaxies is given in Section~\ref{sec:discussion_individual}, while a comparison of clump sizes and SFRs with other literature samples is given in Section~\ref{sec:discussion_literature}.

\begin{table*}
    \begin{tabular}{lrrrrrrllrr}
\hline
    \multicolumn{1}{l}{ID} & \multicolumn{1}{c}{$\rm \mu_{tot}$} & \multicolumn{1}{c}{$\rm \mu_{tan}$} & \multicolumn{1}{c}{$\rm R_{eff}$} & \multicolumn{1}{c}{$\rm Mag_{UV}$}  & \multicolumn{1}{c}{$\rm log(\Sigma_{SFR_{UV}})$} & \multicolumn{1}{c}{$\rm log(M_\star)$} & \multicolumn{1}{c}{$\rm Age$} & \multicolumn{1}{c}{E(B-V)} & \multicolumn{1}{c}{$\rm log\langle\Sigma_{M_\star}\rangle$} & \multicolumn{1}{c}{$\rm T_{cr}$} \\
    \ & \ & \ & \multicolumn{1}{c}{[pc]} & \multicolumn{1}{c}{[AB]} & \multicolumn{1}{c}{[$\rm M_\odot yr^{-1} kpc^{-2}$]}  & \multicolumn{1}{c}{[$\rm M_\odot$]} & \multicolumn{1}{c}{[Myr]} & \multicolumn{1}{c}{[mag]} & \multicolumn{1}{c}{[$\rm M_\odot pc^{-2}$]} & \multicolumn{1}{c}{[Myr]} \\
    \multicolumn{1}{c}{(1)} & \multicolumn{1}{c}{(2)} & \multicolumn{1}{c}{(3)} & \multicolumn{1}{c}{(4)} & \multicolumn{1}{c}{(5)} & \multicolumn{1}{c}{(6)}  & \multicolumn{1}{c}{(7)} & \multicolumn{1}{c}{(8)} & \multicolumn{1}{c}{(9)} & \multicolumn{1}{c}{(10)} & \multicolumn{1}{c}{(11)} \\
\hline
\hline
\multicolumn{9}{l}{MS1358} \\
W\_1	& $7.9^{+7.1}_{-0.9}$	& $-$ 	& $74^{+9}_{-34}$   	& $-20.1^{+1.0}_{-0.1}$	& $2.1^{+0.4}_{-0.4}$	& $9.1^{+1.1}_{-1.1}$	& $1-300$	& $0.0-0.4$	& $4.6^{+1.2}_{-1.1}$	 & $3^{+1}_{-3}$	\\
W\_2	& $16.7^{+10.1}_{-2.8}$	& $-$ 	& $48^{+6}_{-15}$   	& $-18.3^{+0.7}_{-0.2}$	& $1.8^{+0.3}_{-0.3}$	& $8.0^{+0.4}_{-0.8}$	& $1-13$	& $0.2-0.4$	& $3.8^{+0.5}_{-0.8}$	 & $5^{+2}_{-4}$	\\
W\_3	& $8.1^{+7.1}_{-0.9}$	& $-$ 	& $138^{+39}_{-71}$   	& $-17.9^{+1.0}_{-0.2}$	& $0.7^{+0.5}_{-0.5}$	& $9.1^{+0.4}_{-1.4}$	& $1-300$	& $0.0-0.4$	& $4.0^{+0.6}_{-1.4}$	 & $7^{+4}_{-7}$	\\
W\_4	& $9.3^{+6.8}_{-1.2}$	& $-$ 	& $67^{+11}_{-27}$   	& $-17.7^{+0.8}_{-0.1}$	& $1.3^{+0.4}_{-0.4}$	& $7.4^{+0.5}_{-0.9}$	& $1-12$	& $0.1-0.3$	& $2.9^{+0.6}_{-0.9}$	 & $17^{+7}_{-17}$	\\
W\_5	& $12.8^{+8.9}_{-1.9}$	& $6.0$ 	& $66^{+19}_{-49}$   	& $-17.4^{+0.8}_{-0.2}$	& $1.2^{+0.6}_{-0.4}$	& $8.8^{+0.3}_{-1.3}$	& $2-300$	& $0.0-0.4$	& $4.4^{+0.7}_{-1.3}$	 & $3^{+2}_{-3}$	\\
W\_6	& $18.8^{+11.1}_{-3.5}$	& $-$ 	& $109^{+37}_{-48}$   	& $-17.0^{+0.7}_{-0.2}$	& $0.6^{+0.4}_{-0.4}$	& $7.4^{+0.9}_{-1.1}$	& $1-100$	& $0.0-0.4$	& $2.5^{+0.9}_{-1.2}$	 & $34^{+23}_{-34}$	\\
W\_7	& $11.4^{+7.9}_{-1.6}$	& $-$ 	& $118^{+26}_{-48}$   	& $-17.3^{+0.8}_{-0.2}$	& $0.6^{+0.4}_{-0.4}$	& $7.5^{+0.3}_{-0.8}$	& $1-9$	& $0.2-0.3$	& $2.5^{+0.4}_{-0.8}$	 & $36^{+15}_{-24}$	\\
W\_8	& $10.5^{+7.5}_{-1.4}$	& $5.5$ 	& $50^{+45}_{-50}$   	& $-16.2^{+0.8}_{-0.2}$	& $0.9^{+0.9}_{-0.8}$	& $7.7^{+0.6}_{-1.0}$	& $1-90$	& $0.0-0.5$	& $3.5^{+1.1}_{-1.2}$	 & $7^{+10}_{-7}$	\\
W\_9	& $9.6^{+7.6}_{-1.2}$	& $-$ 	& $172^{+54}_{-87}$   	& $-17.6^{+0.9}_{-0.3}$	& $0.4^{+0.5}_{-0.5}$	& $7.5^{+1.3}_{-1.1}$	& $1-200$	& $0.0-0.4$	& $2.2^{+1.4}_{-1.1}$	 & $61^{+37}_{-61}$	\\
W\_10	& $10.0^{+7.8}_{-1.3}$	& $4.9$ 	& $125^{+62}_{-114}$   	& $-17.2^{+0.9}_{-0.2}$	& $0.5^{+0.8}_{-0.6}$	& $8.9^{+0.7}_{-1.5}$	& $1-400$	& $0.0-0.6$	& $4.0^{+1.0}_{-1.6}$	 & $7^{+6}_{-7}$	\\
E\_1	& $4.2^{+1.4}_{-0.8}$	& $-$ 	& $101^{+18}_{-23}$   	& $-19.8^{+0.4}_{-0.2}$	& $1.8^{+0.2}_{-0.2}$	& $8.9^{+0.9}_{-0.6}$	& $1-200$	& $0.0-0.4$	& $4.1^{+1.0}_{-0.6}$	 & $5^{+2}_{-5}$	\\
E\_2	& $10.8^{+5.1}_{-1.3}$	& $7.0$ 	& $34^{+9}_{-18}$   	& $-18.3^{+0.5}_{-0.1}$	& $2.1^{+0.5}_{-0.3}$	& $7.9^{+0.5}_{-0.7}$	& $1-14$	& $0.2-0.4$	& $4.1^{+0.7}_{-0.7}$	 & $3^{+2}_{-3}$	\\
E\_4	& $7.3^{+2.7}_{-0.9}$	& $5.0$ 	& $57^{+31}_{-36}$   	& $-17.3^{+0.4}_{-0.2}$	& $1.3^{+0.6}_{-0.5}$	& $7.7^{+0.4}_{-0.6}$	& $1-11$	& $0.2-0.4$	& $3.4^{+0.7}_{-0.8}$	 & $9^{+8}_{-9}$	\\
E\_5	& $7.2^{+3.2}_{-1.2}$	& $-$ 	& $130^{+37}_{-46}$   	& $-17.6^{+0.5}_{-0.2}$	& $0.7^{+0.3}_{-0.3}$	& $8.5^{+0.3}_{-1.0}$	& $4-100$	& $0.0-0.3$	& $3.5^{+0.4}_{-1.0}$	 & $12^{+7}_{-8}$	\\
E\_6	& $12.7^{+5.9}_{-1.4}$	& $8.3$ 	& $73^{+38}_{-50}$   	& $-16.6^{+0.5}_{-0.2}$	& $0.8^{+0.6}_{-0.5}$	& $7.3^{+0.9}_{-0.7}$	& $1-100$	& $0.0-0.4$	& $2.8^{+1.1}_{-0.8}$	 & $20^{+17}_{-20}$	\\
E\_7	& $7.8^{+3.2}_{-1.0}$	& $-$ 	& $109^{+33}_{-39}$   	& $-17.2^{+0.5}_{-0.2}$	& $0.7^{+0.3}_{-0.3}$	& $6.9^{+0.7}_{-0.7}$	& $1-13$	& $0.0-0.3$	& $2.0^{+0.8}_{-0.7}$	 & $61^{+35}_{-61}$	\\
E\_9	& $5.4^{+2.1}_{-0.9}$	& $-$ 	& $113^{+48}_{-52}$   	& $-17.3^{+0.5}_{-0.3}$	& $0.7^{+0.4}_{-0.4}$	& $7.0^{+1.8}_{-0.6}$	& $1-100$	& $0.0-0.3$	& $2.1^{+1.9}_{-0.7}$	 & $56^{+40}_{-56}$	\\
E\_10	& $6.1^{+2.4}_{-1.0}$	& $4.0$ 	& $151^{+109}_{-122}$   	& $-17.3^{+0.5}_{-0.3}$	& $0.4^{+0.7}_{-0.7}$	& $8.2^{+0.8}_{-0.7}$	& $1-200$	& $0.0-0.5$	& $3.0^{+1.1}_{-0.9}$	 & $22^{+25}_{-22}$	\\
N\_1	& $3.6^{+1.5}_{-1.0}$	& $-$ 	& $121^{+22}_{-29}$   	& $-19.9^{+0.5}_{-0.3}$	& $1.7^{+0.2}_{-0.2}$	& $8.4^{+1.4}_{-1.0}$	& $1-100$	& $0.0-0.4$	& $3.4^{+1.4}_{-1.0}$	 & $13^{+7}_{-13}$	\\
\hline
\multicolumn{9}{l}{RCS0224} \\
E\_1	& $28.2^{+19.4}_{-2.7}$	& $16.7$ 	& $16^{+3}_{-11}$   	& $-17.9^{+0.7}_{-0.1}$	& $2.6^{+0.6}_{-0.3}$	& $7.5^{+0.2}_{-0.8}$	& $1-20$	& $0.0-0.2$	& $4.3^{+0.6}_{-0.9}$	 & $2^{+1}_{-2}$	\\
E\_2	& $47.3^{+55.2}_{-8.0}$	& $27.1$ 	& $17^{+3}_{-17}$   	& $-16.3^{+1.3}_{-0.2}$	& $1.9^{+0.9}_{-0.5}$	& $7.2^{+0.7}_{-1.4}$	& $1-70$	& $0.0-0.3$	& $4.0^{+1.1}_{-1.4}$	 & $3^{+1}_{-3}$	\\
E\_3	& $52.2^{+63.4}_{-12.0}$	& $29.5$ 	& $25^{+6}_{-25}$   	& $-16.6^{+1.3}_{-0.3}$	& $1.7^{+0.9}_{-0.6}$	& $6.6^{+0.3}_{-1.3}$	& $11-15$	& $0.0-0.0$	& $3.0^{+0.9}_{-1.3}$	 & $10^{+3}_{-10}$	\\
W\_2	& $35.3^{+10.4}_{-6.8}$	& $19.7$ 	& $<9^{+2}_{-3}$   	& $-16.4^{+0.3}_{-0.2}$	& $2.5^{+0.3}_{-0.2}$	& $7.4^{+0.7}_{-0.4}$	& $1-90$	& $0.0-0.3$	& $>4.7^{+0.8}_{-0.5}$	 & $<1^{+0}_{-1}$	\\
W\_3	& $36.7^{+10.6}_{-8.1}$	& $20.3$ 	& $24^{+5}_{-7}$   	& $-16.7^{+0.3}_{-0.2}$	& $1.7^{+0.3}_{-0.2}$	& $6.3^{+0.4}_{-0.3}$	& $1-12$	& $0.0-0.1$	& $2.8^{+0.5}_{-0.4}$	 & $12^{+4}_{-9}$	\\
C\_2,3	& $82.4^{+21.9}_{-25.7}$	& $44.8$ 	& $6^{+3}_{-3}$   	& $-16.0^{+0.3}_{-0.3}$	& $2.7^{+0.4}_{-0.4}$	& $7.0^{+0.7}_{-0.6}$	& $1-60$	& $0.0-0.2$	& $4.7^{+0.8}_{-0.8}$	 & $1^{+1}_{-1}$	\\
\hline
\multicolumn{9}{l}{MACS0940} \\
SE\_1	& $26.1^{+0.8}_{-19.6}$	& $14.3$ 	& $46^{+37}_{-14}$   	& $-17.0^{+0.1}_{-0.8}$	& $1.3^{+0.4}_{-0.7}$	& $7.1^{+0.8}_{-0.1}$	& $1-2$	& $0.1-0.2$	& $3.0^{+0.9}_{-0.7}$	 & $13^{+15}_{-6}$	\\
SE\_2	& $25.3^{+0.7}_{-18.9}$	& $13.9$ 	& $24^{+18}_{-1}$   	& $-17.4^{+0.0}_{-0.8}$	& $2.0^{+0.3}_{-0.6}$	& $7.9^{+0.9}_{-0.5}$	& $3-90$	& $0.0-0.3$	& $4.3^{+0.9}_{-0.8}$	 & $2^{+2}_{-2}$	\\
SE\_3	& $25.6^{+0.7}_{-19.1}$	& $14.0$ 	& $<14^{+11}_{-0}$   	& $-17.6^{+0.0}_{-0.8}$	& $2.5^{+0.3}_{-0.6}$	& $7.5^{+1.0}_{-0.1}$	& $11-40$	& $0.0-0.1$	& $>4.4^{+1.0}_{-0.7}$	 & $<2^{+2}_{-2}$	\\
SE\_4	& $25.6^{+0.9}_{-19.1}$	& $13.9$ 	& $55^{+41}_{-2}$   	& $-17.3^{+0.0}_{-0.8}$	& $1.2^{+0.3}_{-0.6}$	& $7.7^{+0.8}_{-0.5}$	& $11-40$	& $0.0-0.1$	& $3.5^{+0.8}_{-0.8}$	 & $8^{+10}_{-1}$	\\
SE\_5	& $27.1^{+0.8}_{-20.5}$	& $14.5$ 	& $<14^{+146}_{-14}$   	& $-15.3^{+0.2}_{-0.8}$	& $1.6^{+0.9}_{-9.2}$	& $-$	& $-$	& $-$	& $-$	 & $-$	\\
SE\_6	& $29.2^{+1.0}_{-22.5}$	& $15.4$ 	& $<14^{+19}_{-14}$   	& $-14.8^{+0.3}_{-0.9}$	& $1.4^{+0.9}_{-1.2}$	& $6.2^{+1.0}_{-0.5}$	& $1-30$	& $0.0-0.2$	& $>3.1^{+1.3}_{-1.3}$	 & $<6^{+13}_{-6}$	\\
SE\_7	& $33.5^{+1.5}_{-26.6}$	& $17.1$ 	& $44^{+45}_{-29}$   	& $-15.7^{+0.2}_{-0.9}$	& $0.8^{+0.7}_{-0.9}$	& $6.6^{+0.9}_{-0.4}$	& $1-15$	& $0.0-0.2$	& $2.5^{+1.1}_{-1.0}$	 & $22^{+35}_{-22}$	\\
SE\_8	& $26.1^{+0.9}_{-19.6}$	& $14.1$ 	& $49^{+36}_{-2}$   	& $-16.3^{+0.0}_{-0.8}$	& $0.9^{+0.3}_{-0.7}$	& $-$	& $-$	& $-$	& $-$	 & $-$	\\
SW\_1	& $13.2^{+15.7}_{-2.7}$	& $-$ 	& $127^{+13}_{-75}$   	& $-17.9^{+1.3}_{-0.2}$	& $0.7^{+0.5}_{-0.5}$	& $7.4^{+0.2}_{-1.3}$	& $1-10$	& $0.1-0.1$	& $2.4^{+0.6}_{-1.3}$	 & $44^{+9}_{-39}$	\\
SW\_2	& $14.0^{+15.3}_{-2.9}$	& $9.9$ 	& $<20^{+4}_{-20}$   	& $-17.6^{+1.2}_{-0.2}$	& $2.2^{+0.9}_{-0.5}$	& $7.4^{+0.2}_{-1.2}$	& $3-13$	& $0.1-0.2$	& $>4.0^{+0.9}_{-1.2}$	 & $<3^{+1}_{-3}$	\\
SW\_3	& $14.2^{+13.3}_{-3.0}$	& $10.0$ 	& $39^{+8}_{-37}$   	& $-18.2^{+1.0}_{-0.2}$	& $1.9^{+0.8}_{-0.4}$	& $8.1^{+0.2}_{-1.2}$	& $11-40$	& $0.0-0.1$	& $4.1^{+0.8}_{-1.2}$	 & $3^{+2}_{-3}$	\\
SW\_4	& $13.8^{+8.7}_{-3.5}$	& $9.8$ 	& $<21^{+12}_{-17}$   	& $-17.2^{+0.7}_{-0.3}$	& $2.0^{+0.7}_{-0.6}$	& $7.6^{+0.3}_{-0.7}$	& $1-11$	& $0.2-0.3$	& $>4.2^{+0.8}_{-0.9}$	 & $<2^{+2}_{-2}$	\\
SW\_5,6	& $11.8^{+2.4}_{-4.1}$	& $8.4$ 	& $29^{+22}_{-21}$   	& $-16.4^{+0.3}_{-0.4}$	& $1.4^{+0.7}_{-0.7}$	& $7.2^{+0.7}_{-0.3}$	& $1-40$	& $0.0-0.3$	& $3.5^{+1.0}_{-0.8}$	 & $6^{+7}_{-6}$	\\
SW\_7	& $11.6^{+1.9}_{-4.4}$	& $8.4$ 	& $30^{+38}_{-30}$   	& $-16.1^{+0.3}_{-0.5}$	& $1.3^{+0.9}_{-1.1}$	& $6.7^{+0.9}_{-0.4}$	& $1-30$	& $0.0-0.2$	& $3.0^{+1.2}_{-1.2}$	 & $11^{+21}_{-11}$	\\
N\_9	& $12.8^{+0.1}_{-8.7}$	& $7.2$ 	& $40^{+42}_{-32}$   	& $-16.5^{+0.1}_{-0.7}$	& $1.2^{+0.8}_{-0.9}$	& $-$	& $-$	& $-$	& $-$	 & $-$	\\
N\_10	& $12.9^{+0.2}_{-8.7}$	& $7.2$ 	& $110^{+92}_{-56}$   	& $-16.4^{+0.2}_{-0.7}$	& $0.2^{+0.5}_{-0.7}$	& $-$	& $-$	& $-$	& $-$	 & $-$	\\
N\_11	& $12.5^{+0.2}_{-8.3}$	& $7.0$ 	& $80^{+72}_{-49}$   	& $-16.5^{+0.2}_{-0.7}$	& $0.5^{+0.6}_{-0.8}$	& $6.3^{+0.7}_{-0.1}$	& $1-10$	& $0.0-0.0$	& $1.7^{+0.9}_{-0.8}$	 & $72^{+99}_{-70}$	\\
E\_9	& $6.5^{+1.5}_{-1.8}$	& $4.3$ 	& $58^{+122}_{-58}$   	& $-16.8^{+0.3}_{-0.4}$	& $1.0^{+0.9}_{-1.8}$	& $7.1^{+0.7}_{-0.3}$	& $1-50$	& $0.0-0.2$	& $2.8^{+1.1}_{-1.9}$	 & $18^{+58}_{-18}$	\\
E\_10	& $5.8^{+1.1}_{-1.7}$	& $3.8$ 	& $81^{+24}_{-15}$   	& $-16.5^{+0.2}_{-0.3}$	& $0.6^{+0.2}_{-0.3}$	& $-$	& $-$	& $-$	& $-$	 & $-$	\\
E\_11	& $5.7^{+1.0}_{-1.7}$	& $-$ 	& $197^{+30}_{-17}$   	& $-17.5^{+0.2}_{-0.3}$	& $0.2^{+0.2}_{-0.2}$	& $-$	& $-$	& $-$	& $-$	 & $-$	\\
\hline
    \end{tabular}
    \caption{Main clump properties: (1) clump ID; (2) total magnification; (3) tangential magnification, reported only if used to derive the intrinsic size, $\rm R_{eff}$; (4) intrinsic effective radius; (5) intrinsic absolute UV magnitude; (6) SFR (UV-derived) surface density; (7) stellar mass; (8) age range of uncertainty; (9) range of uncertainty for the color excess; (10) mass surface density; (11) crossing time, defined by eq.~\ref{eq:Tcr}.}
    \label{tab:results}
\end{table*}

\subsection{Clumps masses}
\label{sec:results_sed}
We report in Tab.~\ref{tab:results} the properties derived from the broadband SED fitting of the clumps. Due to the large uncertainties on ages and extinctions, we choose to provide only a range of allowed values for those two degenerate properties. 
The majority of clumps are consistent with being as young as $\sim1$ Myr (this is consistent with the observations of bright nebular emission in Ly$\alpha$, \CIV\ and [\OII] reported in literature, e.g. \citealp{smit2017,claeyssens2019}) but with equally-probable solutions at ages as old as $\sim100$ Myr. In the latter case, a star formation longer than what is considered should be accounted for, in order to explain the nebular emission. The derived color excesses mainly span the range E(B-V)=0.0-0.3 mag (up to $0.6$ mag in MS1358), suggesting low overall extinctions in these galaxies. 

Model degeneracies have lower effect on clump masses, which we provide as best-fit values; intrinsic masses are affected by uncertainties on the lens model, as seen for absolute magnitudes in the previous section.
Derived clump masses span the range $\rm M=10^6-10^9\ M_\odot$.
Using models with lower metallicity ($\rm Z=0.02\ Z_\odot$) than the reference one ($\rm Z=0.2\ Z_\odot$) we get, on average, the same masses (within 0.1 dex) as the ones reported in Tab.~\ref{tab:results} for MS1358 and MACS0940. In the case of RCS0224, we get 0.3 dex larger masses associated to older best-fit ages, on average. 
We also expect that, using models with longer SFHs, we would derive, on average, larger masses (e.g. by $\sim0.1$ dex for 100 Myr continuous star formation, see \citealp{messa2022}). 
Both these alternative assumptions would cause mass differences that are within the mass uncertainty ranges reported in Tab.~\ref{tab:results}.

\begin{figure*}
    \includegraphics[width=0.33\textwidth]{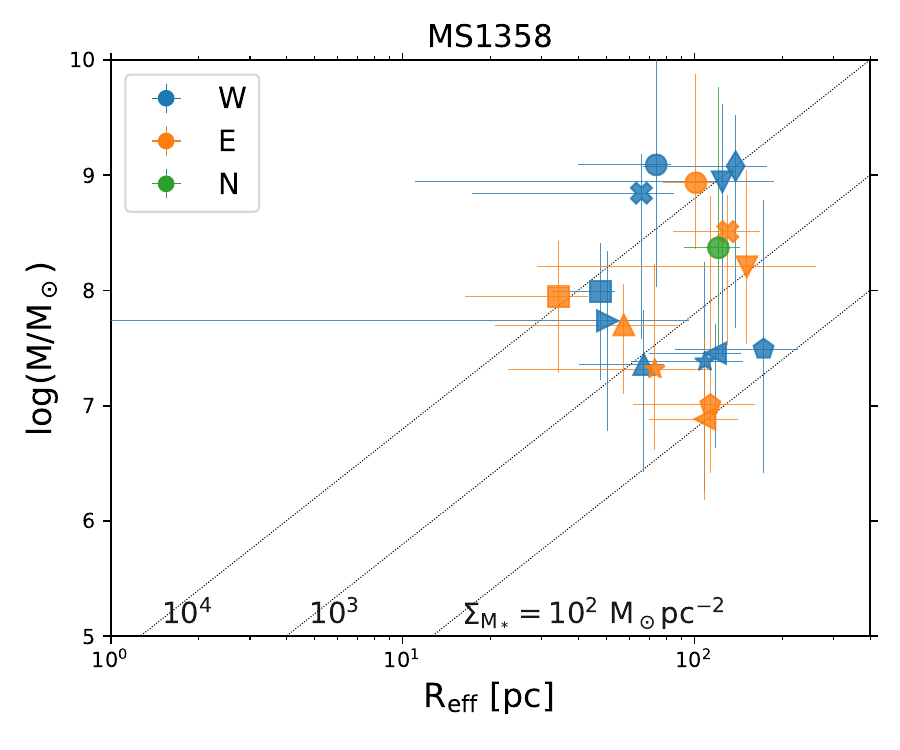} 
    \includegraphics[width=0.33\textwidth]{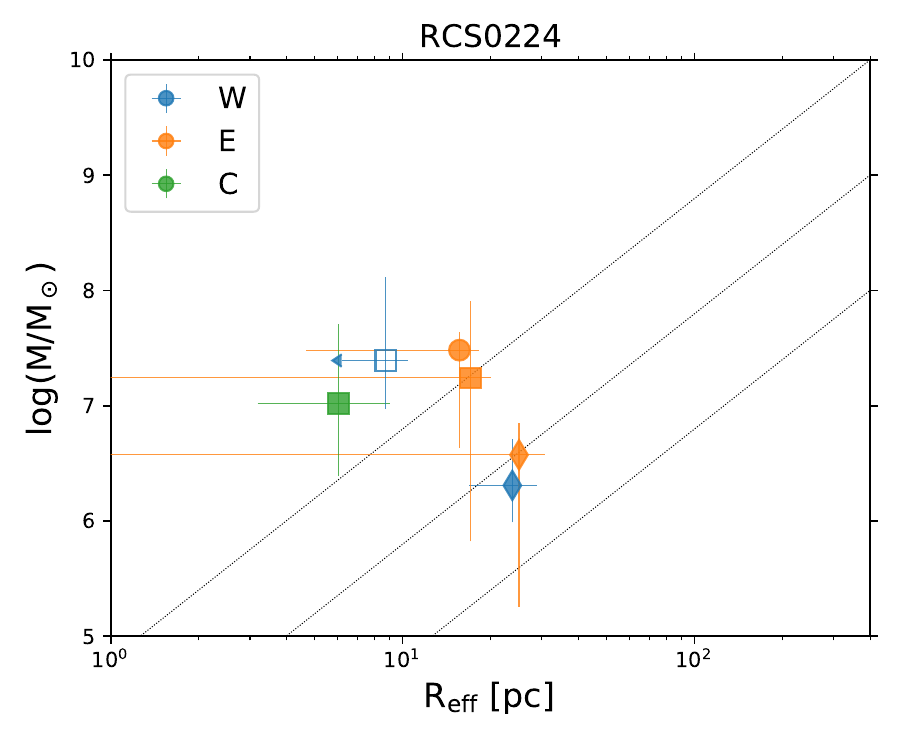}
    \includegraphics[width=0.33\textwidth]{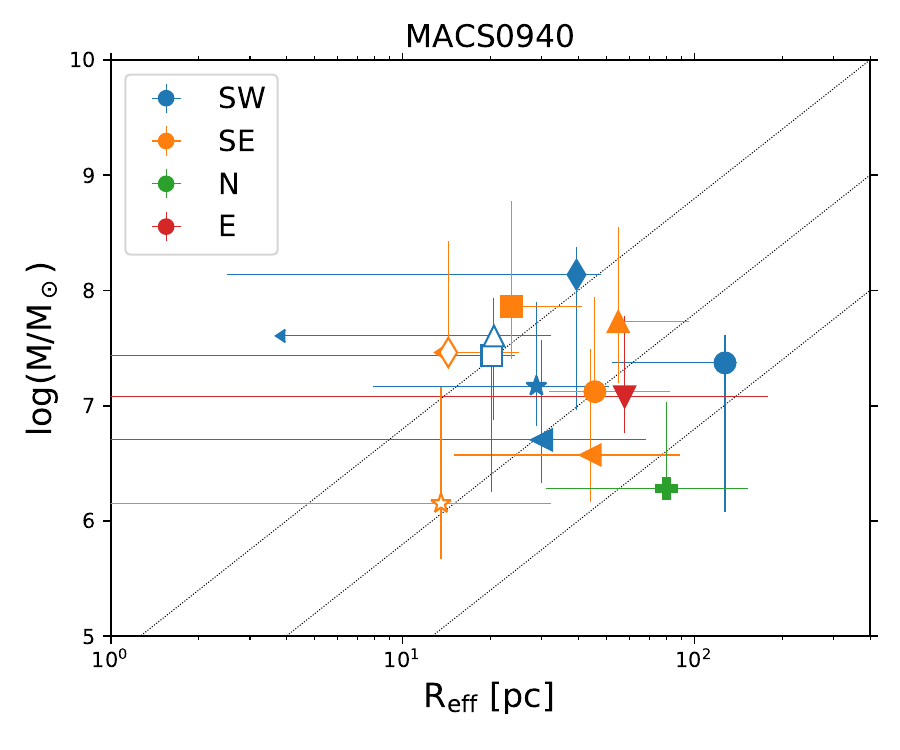}
    \caption{Intrinsic sizes and masses of the clumps. Solid gray lines are of constant mass surface density for $10^2$, $10^3$ and $10^4$ $\rm M_\odot pc^{-2}$. The same color and marker notation as in Fig.~\ref{fig:phot_results} is used.}
    \label{fig:SED_results}
\end{figure*}

The comparison of clump masses to their sizes, in Fig.~\ref{fig:SED_results}, reveals that we are looking at very dense systems; they have, on average, mass surface densities $\rm \Sigma_{M_\star}\sim10^3\ M_\odot pc^{-2}$, similar to those of stellar clusters in local galaxies \citep[e.g.][]{brown2021} but on scales which are up to $\sim10$ times larger, i.e. up to $\sim100$ pc in the case of MS1358. The densest systems observed in these three high-z galaxies reach values $\rm >10^4\ M_\odot pc^{-2}$, matching the most extreme stellar ensembles observed at any redshift (see the discussion in Section~\ref{sec:discussion_literature}). 

The combination of mass ($\rm M_\star$) and size ($\rm R_{eff}$) of the clumps provides an estimate of their crossing time, defined as \citep{gieles2011}:
\begin{equation}
\label{eq:Tcr}
    \rm T_{cr}\equiv10\left(\frac{R^3_{eff}}{GM_\star}\right)^{\frac{1}{2}}
\end{equation}
In local studies of stellar clusters, crossing times are compared to cluster ages to derive the so-called `dynamical age' of clusters ($\rm \Pi\equiv Age/T_{cr}$); a value $\Pi>1$ indicates that the stars in the system remained clustered together, without freely expanding into their surroundings, for a time longer than their crossing time, implying that the system is likely gravitationally bound \citep[e.g.][]{gieles2011,ryon2015,ryon2017,krumholz2019,brown2021}. The same kind of analysis has been recently applied to the study of high-z stellar clumps \citep[e.g.][]{vanzella2021,messa2022,claeyssens2023,vanzella2022_sunrise}. Crossing times for the clumps in the three galaxies of the current sample are reported in Tab.~\ref{tab:results}. Given the large age uncertainties, we do not attempt to calculate the respective dynamical ages; however, we notice that $\rm T_{cr}\lesssim10$ Myr for $\sim55\%$ of the cases, suggesting that (even if their young ages are confirmed) a large fraction of the clumps we are observing can be of gravitationally bound systems. 

\section{Discussion}
\label{sec:discussion}
\subsection{Stellar clump populations}
\label{sec:discussion_individual}
We discuss in this section the results presented in Fig.~\ref{fig:phot_results} and Fig.~\ref{fig:SED_results}, and collected in Tab.~\ref{tab:results}, by putting the clump properties in the context of their host, for each of the galaxies studied. 
\subsubsection{The clump population of MS1358}
MS1358 is characterised by the presence of several clumps; 10 were extracted in the western image (the one with the largest overall magnification), 8 of which are seen also in the eastern image. 
The analyses on the two images of the galaxy, produce comparable clump properties (see Tab.~\ref{tab:results}; Figs.~\ref{fig:phot_results} and \ref{fig:SED_results}). For clump 8, visible only in the western part of the galaxy, we were not able to derive a robust size estimate.  
As suggested by previous studies, and as clearly visible from Fig.~\ref{fig:galaxies1}, MS1358 is dominated, in the rest-frame UV, by two main clumps (ID: 1 and 2); those account for $\sim25\%$ of the flux in F775W, rest-frame $\sim1300$ \AA\ (clump 1 alone accounts for $\sim20\%$, while if all clumps are considered, their contribution to the rest-frame UV emission of the galaxy is $\sim40\%$), suggesting that they are major contributors to the recent star formation of the galaxy. Their rest-frame UV fluxes, converted to the SFR via the \citet{kennicutt2012} relation, correspond to 5 and 1 $\rm M_\odot yr^{-1}$. 
Nebular tracers of the current star formation indicate that their contribution is even larger ($\sim40\%$ of the SFR of the galaxy, \citealp{swinbank2009}). The [\OII]-derived SFR, $7.5$ and $\rm 3.7\ M_\odot yr^{-1}$ for the two clumps\footnote{Adapted from original values $\rm SFR = 12$ and $\rm 6\ M_\odot yr^{-1}$ reported in \citet{swinbank2009}, see footnote~\ref{note1}.}), suggests that the $\rm SFR_{UV}$ values we derived for our clumps may be underestimated; this could either be due to the SFR episode in the galaxy being more recent than what is assumed by the \citet{kennicutt2012} conversion and/or due to the presence of some extinction. 
The derived masses span the range $\rm 10^7-10^9\ M_\odot$ and are therefore comparable to the mass of the entire galaxy as estimated by \citet{swinbank2009}. The derived young ages for some of the clumps ($\lesssim10$ Myr), combined to the nebular emission observed in the galaxy \citep{swinbank2009} suggest that a considerable fraction of the stellar mass in the galaxy is being formed during the current star formation episode, and that the clumps are an important contributor to that process. 

\subsubsection{The clump population of RCS0224}
RCS0224 is characterised by 3 bright clumps, accounting for $\rm 45\%$ of the rest-frame UV emission. The brightest one (ID: 1, $\rm SFR_{UV}=0.7\ M_\odot yr^{-1}$) is not multiply imaged and therefore is only visible in the eastern image of the arc. The other two clumps do not have robust intrinsic size (and flux) estimates in the eastern image, due to the large uncertainties related to the magnification values; we can however rely on their properties derived in the western image. They are seen down to very small scales; both their sizes (below 10 pc for clump 2) and masses ($\rm 10^6-10^7\ M_\odot$) are close to the ones of super star clusters in nearby galaxies \citep[e.g.][]{leitherer2018,vanzella2019,adamo2020}. The mass surface density of clump 3 ($\rm \Sigma_{M_\star}\sim10^3\ M_\odot pc^{-2}$) is also consistent with nearby star clusters \citep[e.g.][]{brown2021}, while clump 2 has a density close to  $\rm \Sigma_{M_\star}\sim10^5\ M_\odot pc^{-2}$, one of the largest observed for star-forming regions (see also the discussion in Section~\ref{sec:discussion_literature}). 
A single source is visible in the central image of the galaxy (see Fig~\ref{fig:galaxies2}) but the lensing model cannot distinguish which source it is.
The intrinsic size of the source is 6 pc and both the luminosity and mass suggest that this is another image of clump 2.

Summing the $\rm SFR_{UV}$ of the clumps in the central and western images and comparing it to the value estimated from [\OII] emission in the same region ($\rm 8.2\ M_\odot yr^{-1}$, \citealp{swinbank2007}), we derive that only $<10\%$ of the galaxy SF is in the observed clumps. This result is consistent with bright [\OII] emission being observed along the entire arc seen in F814W (a long part of which is devoid of compact sources, see Fig~\ref{fig:galaxies2}); it may also indicate, as pointed out for MS1358, that $\rm SFR_{UV}$ values are underestimates and therefore that the bulk of star formation in the galaxy is very recent. This galaxy is also the one with the shallowest data (as discussed in Section~\ref{sec:results_phot}), and therefore there is also the possibility that current observations are missing a population of low surface brightness clumps, currently undetected along the arc.

\subsubsection{The clump population of MACS0940}
Compact sources are seen all along the lensed arcs of MACS0940; the galaxy is dominated by 4 bright clumps (IDs: 1 to 4) contributing to $\sim40\%$ of the rest-frame UV emission of the galaxy (the contribution of all the clumps in the arcs is $\sim50\%$). 
For some of the clumps we were not able to derive robust size estimates. SE\_5 and SE\_6 have large size uncertainties; contrary to the rest of the sample, they both have de-lensed magnitudes that differ largely from their SW counterpart (SW\_5,6), hinting to a possible mis-association. We point out that there are no detected SW clumps with similar photometric properties to the former. Many of the clumps in the SW region (e.g. IDs: 2, 3, 4 and 7) also have large size uncertainties, mainly due to uncertainties in the lens model in that region of the arc; on the other hand, their SE counterparts have robust measurement that help inferring their intrinsic properties and will be used in the further analyses of this work, Section~\ref{sec:discussion_literature}. For some clumps we could not perform SED fitting due to the lack of signal in some of the filters (clumps SE\_5, SE\_7, N\_10, N\_11, E\_11 and E\_12). Sizes and masses of the bright clumps in the galaxy, distributed mainly between $15-60$ pc and $\rm 10^6-10^8\ M_\odot$, lead to large mass surface densities, $\rm \Sigma_{M_\star}=10^3-10^4\ M_\odot pc^{-2}$.
While little is known about the properties of MACS0940, the clump contribution to the rest-frame UV emission and the spatial superposition of the peaks of \lya\ emission \citep{claeyssens2019} on the location of the brightest clumps suggest that the observed clumps are strongly contributing to the host galaxy recent star formation.

\subsection{Comparison to literature samples}
\label{sec:discussion_literature}
\subsubsection{Size and SFR across redshifts}
\label{sec:discussion_literature_sfr}
\begin{figure*}
    \includegraphics[width=0.49\textwidth]{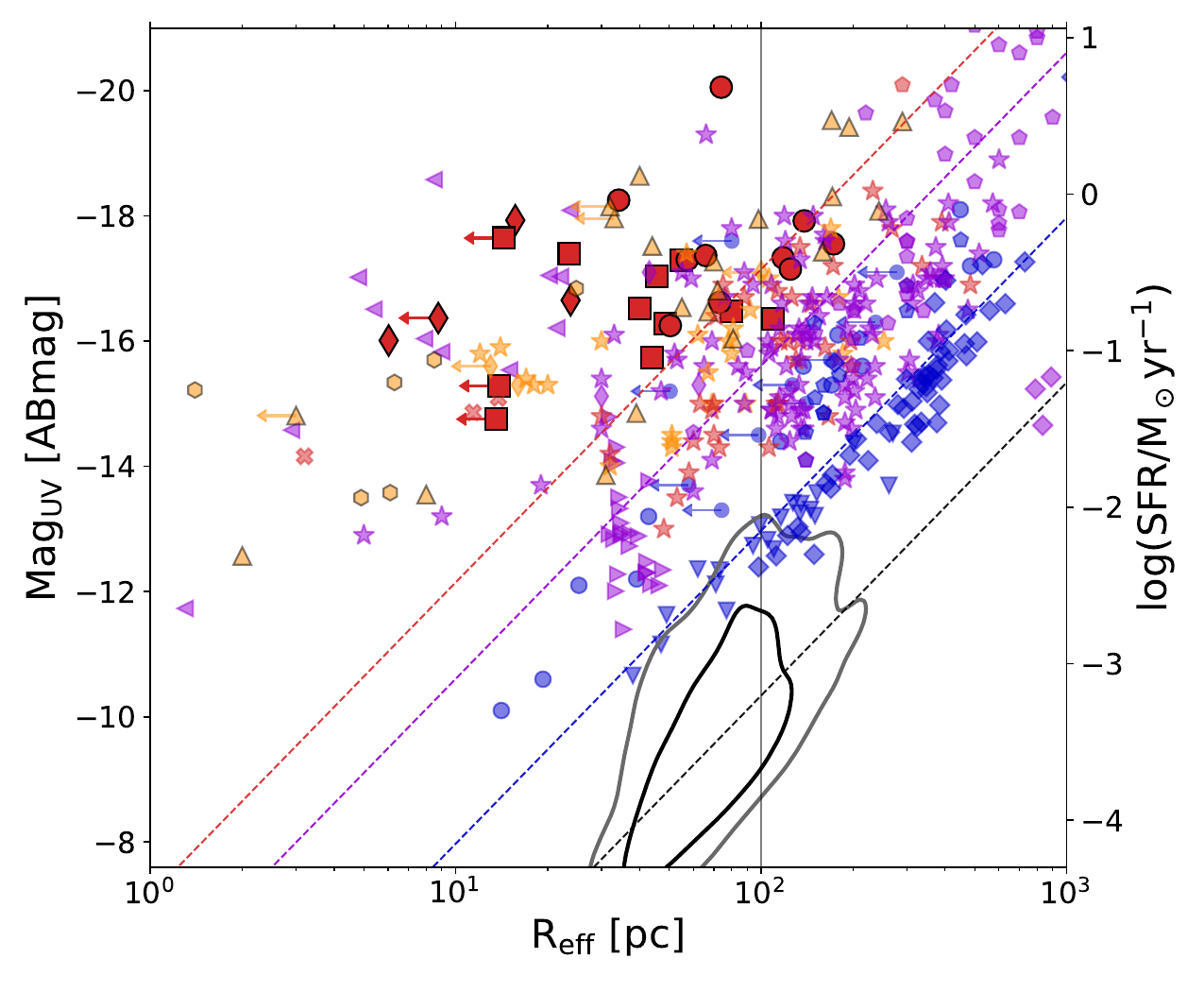}
    \includegraphics[width=0.49\textwidth]{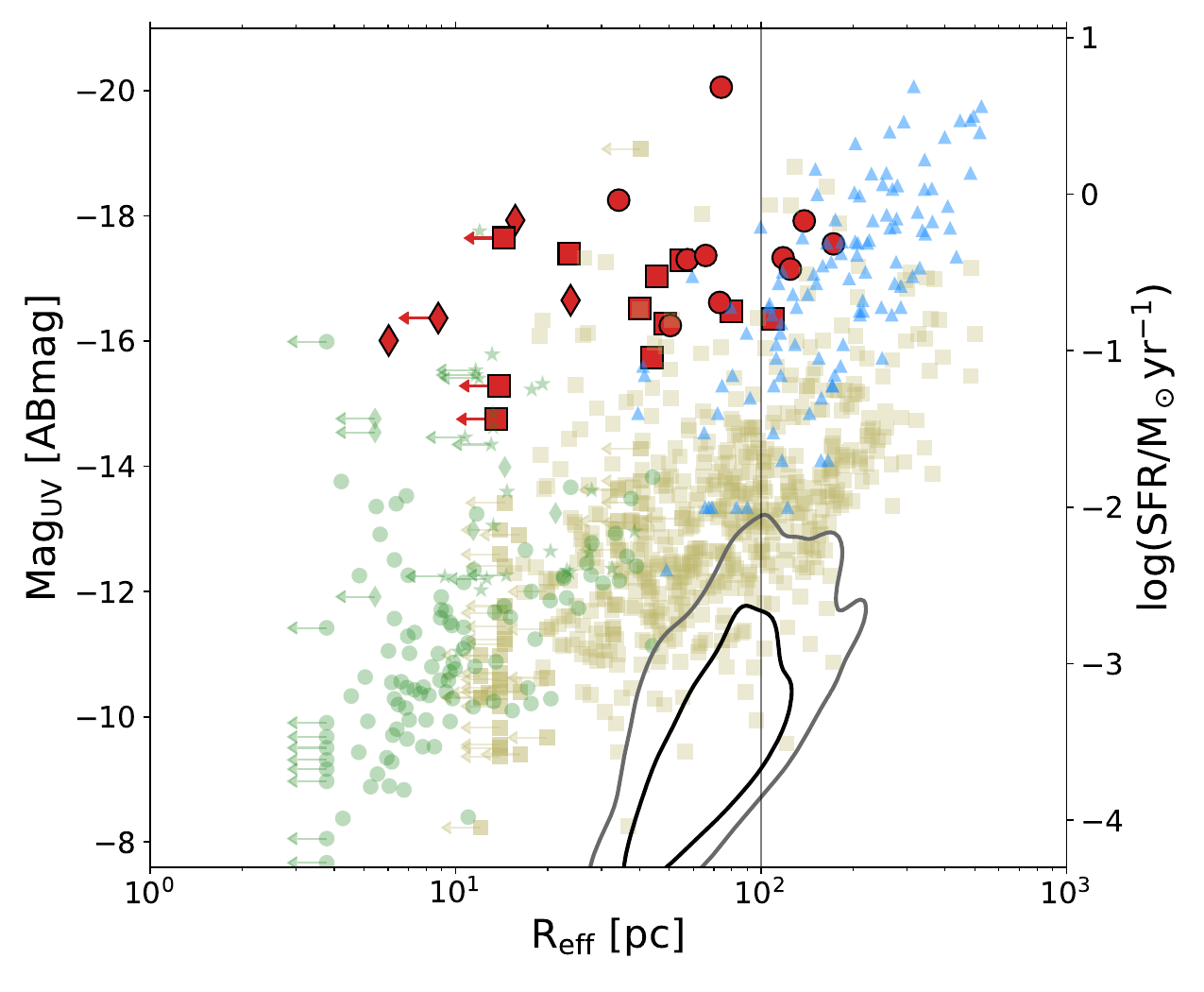}
    \includegraphics[width=0.65\textwidth]{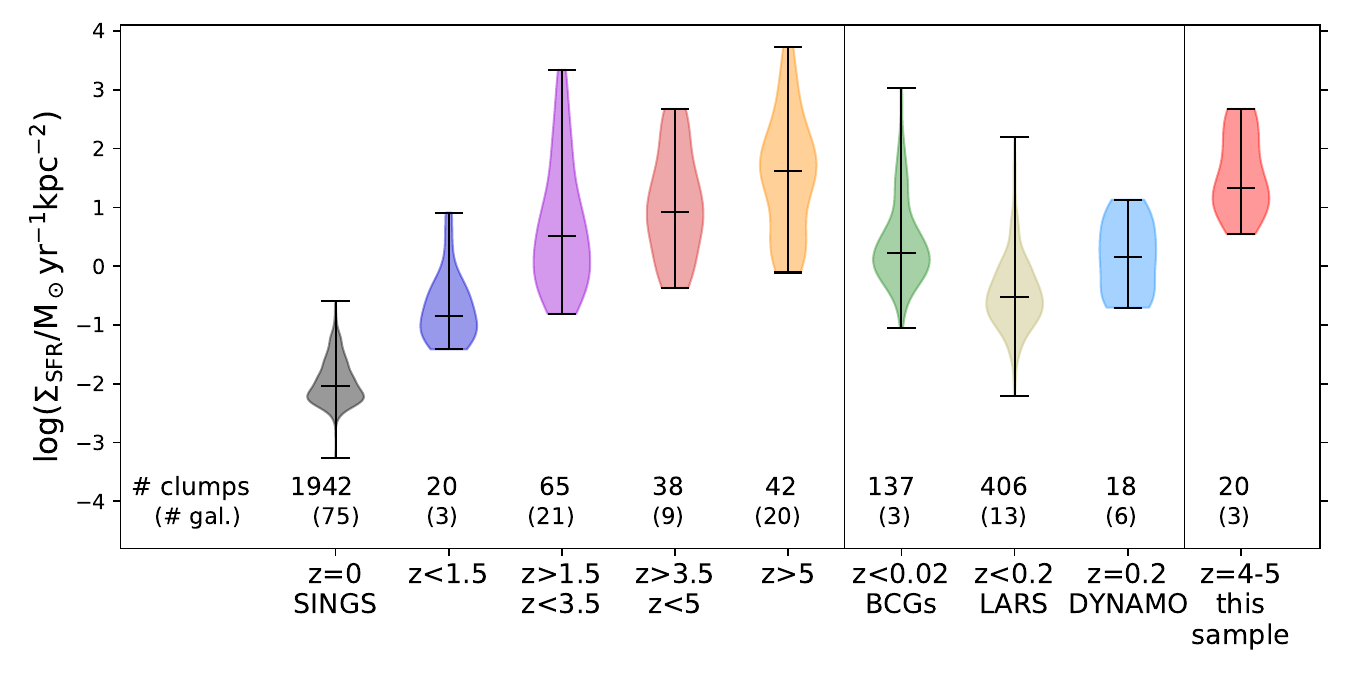}
    \includegraphics[width=0.33\textwidth]{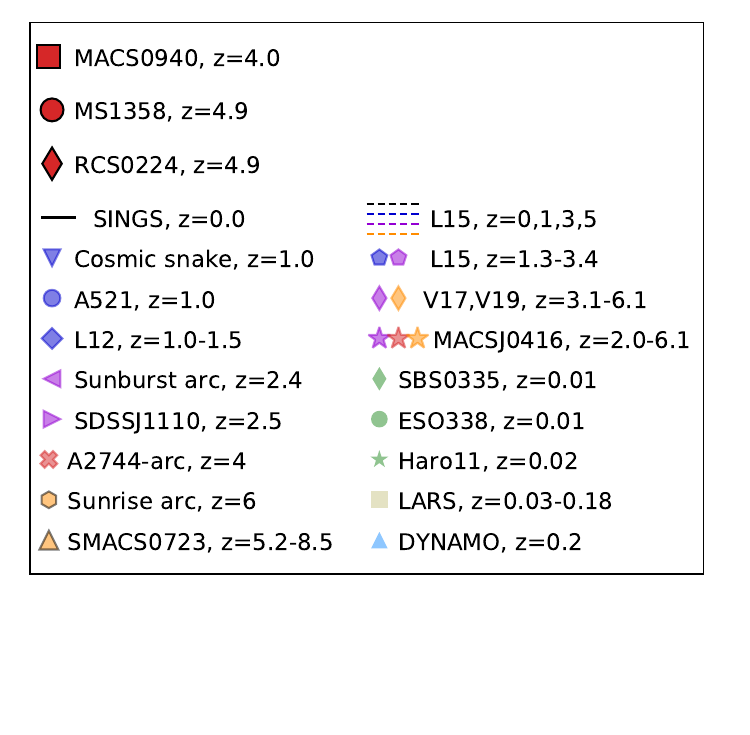}
    \caption{\textit{(Top left:)} sizes and magnitudes of the clumps in the current study compared to clump samples from literature, color-coded by their redshift (blue: $\rm z<1.5$, purple: $\rm 1.5\le z<3.5$, red: $\rm 3.5\le z<5$, orange: $\rm z\ge5$). The sample at $\rm z=0$ from SINGS \citep{kennicutt2003} is shown as black density contours (enclosing 1 and $\rm 2\sigma$ of the distribution). On the y-axis we are showing either \ha\ luminosities converted to SFR values (for L12 and L15) or UV magnitudes (for the rest of the samples). Values of typical SFR surface densities for samples at $z=0,1,3$ and $5$, as derived by L15, are shown as dashed lines.
    Individual redshifts (or ranges) for each of the samples is given in the bottom-right panel. The references for the samples considered are: the \textit{Cosmic snake} arc \citep{cava2018}, A521-sys1 \citep{messa2022}, L12 \citet{livermore2012}, the \textit{Sunburst} arc \citep{vanzella2022}, SDSSJ1110+6459 \citep{johnson2017}, Abell2744-arc \citep{vanzella2022_a2744}, the \textit{Sunrise} arc \citep{vanzella2022_sunrise}, SMACS0723 (considering only clumps at $\rm z>5$, i.e. where JWST-NIRCam traces their rest-frame UV emission, see \citealp{claeyssens2023}), L15 \citep{livermore2015}, V17\&V19 \citet{vanzella2017a,vanzella2017b,vanzella2019}, MACSJ0416 \citep{mestric2022}. In case of multiply-imaged clumps, we consider only the one with the largest magnification.
    \textit{(Top right:)} comparison to nearby clump samples of: blue compact galaxies (ESO338-IG04, \citealt{messa2019}; SBS0335-052E and Haro11, this work), LARS \citep{messa2019} and DYNAMO \citep{fisher2017a}. On the y-axis we are showing \ha\ luminosities converted to SFR values for DYNAMO, and UV magnitudes for the rest of the samples. \textit{(Bottom left:)} distribution of SFR densities (shown as violin plots, with median and extreme values of the distribution marked) of the clump samples, binned by redshift; only clumps with $\rm R_{eff}<100$ pc are considered. The clumps from the current work are shown also as a separate sample.}
    \label{fig:sizelum_literature}
\end{figure*}

The sizes and luminosities of the clumps in these 3 galaxies are compared to samples from literature in Fig.~\ref{fig:sizelum_literature} (top-left panel);
in case of clumps with multiple images we will consider only the values recovered from the one with highest magnification. The samples are color-coded according to their redshift. 
The figure shows that, at any scale between $\sim1$ and $\sim10^3$ pc, clumps become on average increasingly brighter moving to higher redshifts; a similar trend was already suggested by \citet{livermore2015} using samples studied in \ha\ emission. 
We point out that the SFR scale in Fig.~\ref{fig:sizelum_literature} (derived from the \citealp{kennicutt2012} conversion, as already described in Section~\ref{sec:results_phot}), is used to directly compare samples studied in \ha\ (SINGS, L12, L15, DYNAMO, see figure caption) to the others, studied in rest-frame UV. We are not including, in this study, SFR values derived from the analysis of the clump spectral energy distributions\footnote{The published sample from \citet{mestric2022} includes SED-derived SFR values; however, for consistency with the other samples included in Fig.~\ref{fig:sizelum_literature}, we chose to plot only their rest-frame UV magnitudes.}. 
The average clump $\rm SFR_{H\alpha}$ surface densities for $\rm z=0$, 1, 3 and 5 proposed by \citet{livermore2015} are shown as dashed lines in Fig.~\ref{fig:sizelum_literature} (top-left panel), and are consistent with the overall densities of the plotted samples. In the same figure we show, as black-grey contours (enclosing 1 and 2$\sigma$ of the sample), the sizes and SFRs of \HII\ regions in the SINGS sample of local MS galaxies \citep{kennicutt2003}; already at $\rm z=1$, clumps are clearly detached from the region covered by the SINGS sample, as pointed out by previous studies \citep[e.g.][]{livermore2012,livermore2015,messa2022,claeyssens2023}. 

The redshift evolution of clump $\rm \Sigma_{SFR}$ is made explicit in the bottom-left panel of Fig.~\ref{fig:sizelum_literature} via violin distributions. For the whole sample we consider only clumps with $\rm R_{eff}<100$ pc, as we want to focus this analysis on compact star-forming regions; however, similar results would be derived considering clumps at all scales (see Appendix~\ref{sec:app:alternative_plots}).
A clear redshift evolution of $\rm \Sigma_{SFR}$ is observed comparing local clumps ($\rm z=0$) to samples at cosmic `afternoon' ($\rm 0<z<1.5$) and at cosmic noon ($\rm 1.5\leqslant z<3.5$). A shallower evolution is seen at earlier cosmic times: while the distributions are on average shifted towards denser $\rm \Sigma_{SFR}$ at higher redshifts, extreme values can be observed already at $z\sim2$, as it is the case for the clump population of the \textit{Sunburst} galaxy (purple triangles in the top-left panel of Fig.~\ref{fig:sizelum_literature}), studied at scales $<10$ pc \citep{vanzella2022}.

A necessary caveat is that the surface brightness completeness is different for each of the samples considered. Completeness limits are unavailable for many of the samples; it can however be assumed that they become brighter moving to higher redshift. For example, the completeness limits derived in the current work, $\rm 1-10\ M_\odot yr^{-1}kpc^{-2}$, are larger than the densest clumps at $\rm z<1.5$. We argue that the completeness limits, biasing the study of high-z systems towards the most extreme (i.e. densest) sources, may be the main driver of the $\rm \Sigma_{SFR}$ redshift evolution overall, especially of the median values of their distributions. As mentioned in the discussion of individual galaxies (Section~\ref{sec:discussion_individual}) other factors possibly affecting the overall $\rm \Sigma_{SFR}$ distributions are the ages of the clumps and their extinctions.
On the other hand, we expect that the high-density end of each distribution is not affected by the completeness; in this case the evolution of the densest clumps would be real and may reflect the redshift evolution of the properties of their host galaxies. We also remind that there is a great in-homogeneity in how the considered literature samples were selected and analysed. 
Systematic statistical studies of clumps across redshift ranges are needed to truly prove and map the evolution of clumps; forthcoming studies with the JWST will indeed provide this statistics.

We can interpret the clump $\rm \Sigma_{SFR}$ redshift evolution discussed in this section as an evolution of the host galaxy conditions where the clumps form. Main sequence galaxies are characterised by higher SFR densities at higher redshift, with the most extreme change happening during the cosmic noon ($\rm z=1-3.5$, \citealp[e.g.][]{schreiber2015}). In order to test this possibility we consider, in the next section, the properties of clumps in local starburst galaxies, i.e. galaxies which do not fall into the local main sequence but rather present typical properties of higher redshift systems. 

\subsubsection{Size and SFR of nearby samples}
\label{sec:discussion_literature_nearby}
In order to test the effects of galaxy environments on their stellar clump population, we consider local samples of galaxies characterised by properties typical of high-z systems. Galaxies in the DYNAMO sample, at $\rm z=0.2$, were selected to contain gas-rich galaxies with turbulent and marginally stable disks, well representing the star formation conditions at cosmic noon, $\rm z\sim1-3$ \citep{green2014}; their star-forming regions were studied in \ha\ down to $\sim50$ pc resolution \citep{fisher2017a}. In a similar way, the Lyman-Alpha Reference Sample (LARS), at redshift $\rm z=0.03-0.20$, contains galaxies selected to be analogs of $\rm z\sim3$ (Lyman-break and Lyman-emitter) galaxies, characterised by elevated UV luminosities and SFRs \citep{ostlin2014}; their stellar clumps were studied in far-UV emission down to 10 pc scales \citep{messa2019}. Finally, blue compact galaxies (BCGs) are (usually low-mass) star-forming systems with high specific star formation rate \citep[e.g.][]{ostlin2001};
among the best studied ones, SBS 0335-052E, ESO 338-IG04 and Haro 11 are known to host populations of bright young stellar clusters/clumps \citep[e.g.][]{ostlin2001_sbs0335,ostlin2003,adamo2010b,adamo2011,sirressi2022}. Sizes and UV magnitudes of clumps in ESO 338-IG04 were derived in \citet{messa2019}. No size measurements are available in the literature for clumps in SBS 0335-052E and Haro 11; we therefore perform, for the clumps in these galaxies, the same size-luminosity analysis described in Section~\ref{sec:modeling} in the F140LP filter, tracing FUV emission. 

The sizes, UV magnitudes (and SFRs) and $\rm \Sigma_{SFR}$ distributions of the clumps in these nearby samples are shown in Fig.~\ref{fig:sizelum_literature} (top-right and bottom-left panels). In the case of SBS 0335-052E we plot the  6 main \textit{super star clusters} discussed in \citet{adamo2010b}; for Haro 11 we select, from the sample of \citet{sirressi2022}, the brightest clusters, corresponding to UV magnitudes $<23$ mag; for ESO 338-IG04 and LARS we plot all the clumps from the \citet{messa2019} samples with a photometric uncertainty $<0.3$ mag in UV. Clump sizes span a broad range from $4$ pc (individual stellar clusters) to $\sim1$ kpc (large star-forming regions). When considering only the compact ($\rm R_{eff}<100$ pc) clumps, in Fig.~\ref{fig:sizelum_literature} (bottom-left), their median SFR surface densities are similar to the ones of clumps observed in galaxies at $0<z<3.5$; the most extreme cases are as dense as the clumps in the $\rm z\sim5$ of the sample studied in this work. These distributions suggest that the environmental conditions setting the starburst nature of these galaxies (interactions, mergers, elevated gas fraction and turbulence) can indeed drive the formation of clumps with elevated SFR densities.

\subsubsection{Mass surface densities}
\label{sec:discussion_literature_mass}
\begin{figure*}
    \includegraphics[width=0.46\textwidth]{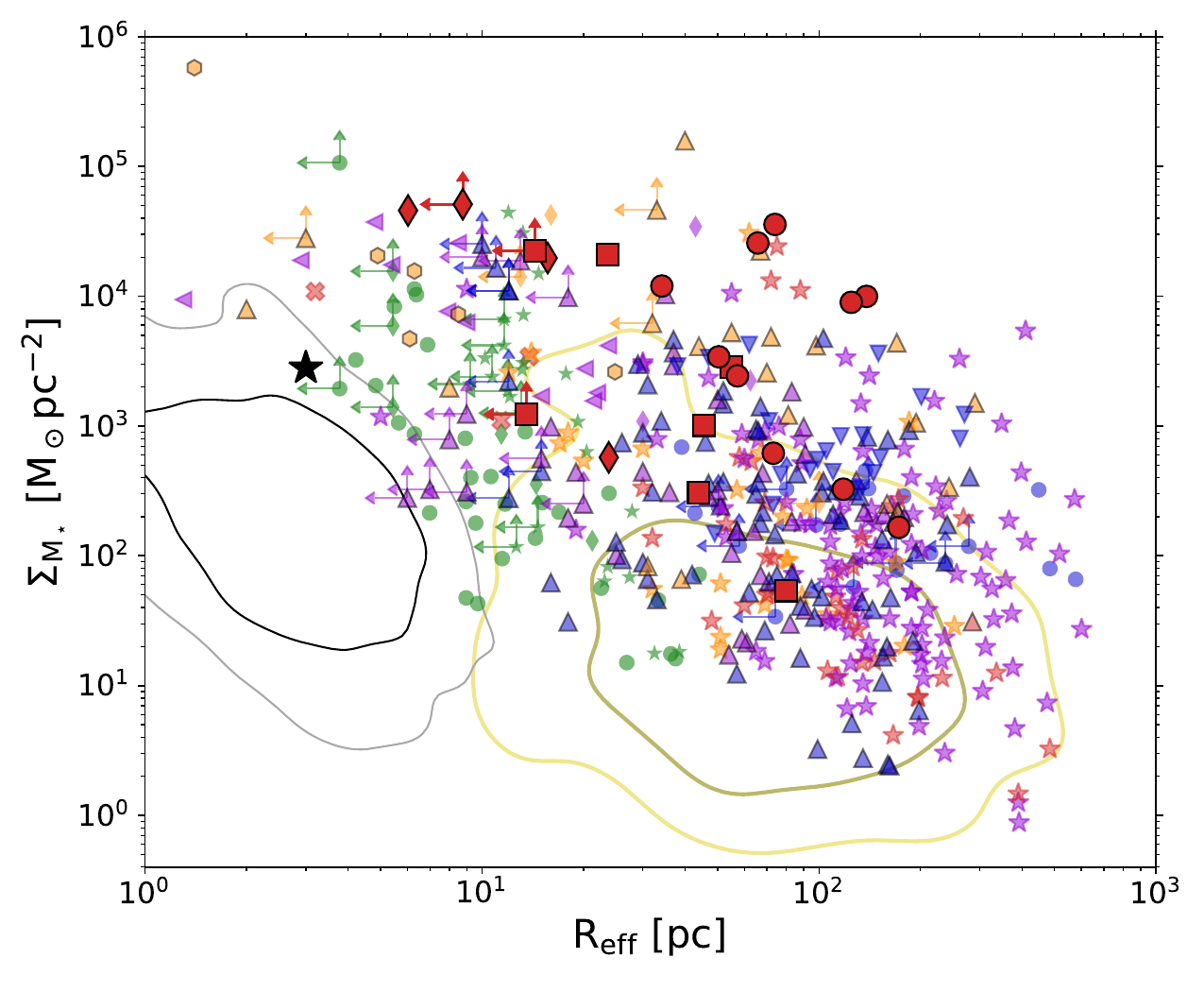}
    \includegraphics[width=0.53\textwidth]{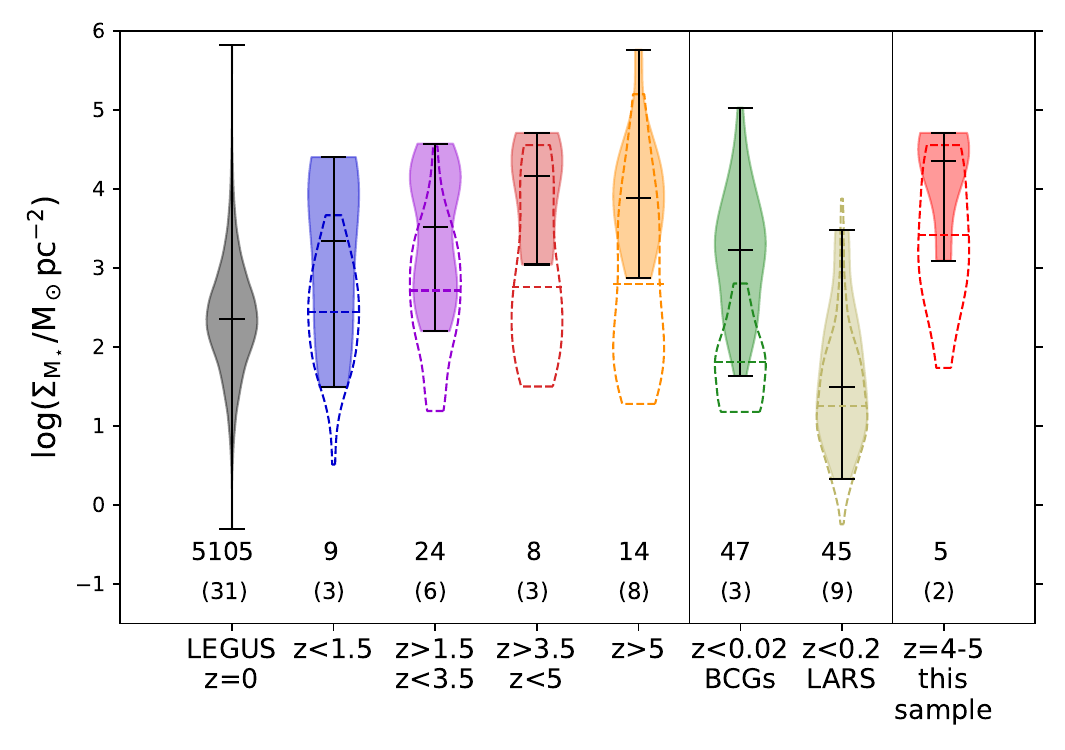}
    \caption{\textit{(left panel)}: mass surface densities of clumps, in function of their intrinsic sizes; the color and marker coding of the plotted samples are the same presented in Fig.~\ref{fig:sizelum_literature}; the only differences are the LARS sample, plotted as yellow contours instead of square markers, and the inclusion of the LEGUS sample of local stellar clusters (from the analysis of \citealp{brown2021}), instead of  SINGS, as black contours, due to the lack of available masses for the latter. Both yellow and black contours enclose 1 and $\rm 2\sigma$ of the relative distribution.
    The location of a typical globular cluster with $\rm M=10^5\ M_\odot$ and $\rm R_{eff}=3 pc$ \citep{brodie2006} is shown with a black star marker. 
    \textit{(right panel)}: $\rm \Sigma_{M_\star}$ distributions, shown as violin plots; the distributions of clumps with $\rm R_{eff}\leqslant20$ pc are shown as filled violins, while clumps in range $\rm 20<R_{eff}\leqslant100$ pc are shown with empty violins with dashed edges (a dashed line marks also the median value of each distribution). 
    }
    \label{fig:sigmaM_literature}
\end{figure*}

We investigate if the SFR redshift evolution discussed in Section~\ref{sec:discussion_literature_sfr} (and Fig.~\ref{fig:sizelum_literature}) is also reflected in the clump masses.
We plot in Fig.~\ref{fig:sigmaM_literature} the clumps mass surface densities,  $\rm \Sigma_{M_\star}$, as a function of the clump sizes (left panel) and as violin distributions (right panel); we use the same clump samples as in Fig.~\ref{fig:sizelum_literature}, when mass values are available. In more detail, masses are unavailable for L12, L15,  SDSSJ1110, the nearby DYNAMO sample and the SINGS sample of local galaxies; instead, we plot the LEGUS sample of local clusters (from the analysis of \citealp{brown2021}), to help the comparison of compact clumps to single stellar clusters. For comparison we provide also the stellar surface density of a typical $\rm 10^{5.2}\ M_\odot$ globular cluster with $\rm R_{eff}=3$ pc \citep{brodie2006}. In this analysis we consider the entire SMACS0723 sample by \citet{claeyssens2023} and not only the one at $\rm z>5$ as done in Fig.~\ref{fig:sizelum_literature}, because the selection here is based on clump mass and not on FUV luminosity.
The references for the samples are the same reported in the caption of Fig.~\ref{fig:sizelum_literature}, except for masses of clumps in: (i) SBS0335-052E, from \citet{adamo2010b}; (ii) Haro 11, from \citet{sirressi2022}; (iii) ESO338-IG04 and LARS, from a forthcoming paper (Messa et al., in prep).

Mass surface densities span almost 5 orders of magnitudes ($\rm \Sigma_{M_\star}\sim10^0-10^5\ M_\odot pc^{-2}$). Contrary to the $\rm \Sigma_{SFR}$ values (Fig.~\ref{fig:sizelum_literature}), no clear redshift trend is visible in the left panel of Fig.~\ref{fig:sigmaM_literature}. Some of the clumps in the current study, from MS1358 and MACS0940, stands out, together with other few $\rm z>1.5$ sources as very dense ($\rm \Sigma_{M_\star}>10^4\ M_\odot pc^{-2}$) large ($20-100$ pc) clumps;
these are the densities of the densest gravitationally bound local stellar systems (young and globular clusters), yet on much larger scales. The absence of a clear redshift evolution of $\rm \Sigma_{M_\star}$ can be deduced also from redshift-binned violin distributions (see Appendix~\ref{sec:app:alternative_plots}). 

A tentative redshift evolution of $\rm \Sigma_{M_\star}$ can be seen when considering the clumps at $\rm R_{eff}\leqslant20$ pc i.e. at the scales of star clusters, Fig.~\ref{fig:sigmaM_literature}, right panel. An apparent feature of the figure is the increase in the upper-end of the $\rm \Sigma_{M_\star}$ distribution at $\rm z>5$, but this is driven by a single massive ($\rm M\sim10^7\ M_\odot$) compact ($\rm R_{eff}=1.4\ pc$) source observed in the \textit{Sunrise} arc \citep{vanzella2022_sunrise}. 
The high stellar densities of high-z clumps become evident when local clusters are considered (black contours and black distribution in Fig.~\ref{fig:sigmaM_literature}); there is only little overlap between local clusters and their counterparts at higher redshifts, the latter being on average denser.
As already suggested in the previous section, this evolution probably reflects the ambient pressure where clusters form; galaxies at higher redshift are characterised by densest environment, which in turn are able to form denser clusters. This environmental effect was observed locally for stellar clusters \citep[e.g.][]{johnson2017,messa2018b,adamo2020}. This trend seems to be confirmed by the clumps in the local starburst BCGs, reaching densities comparable to their $\rm z>1.5$ counterparts.
We would like to point out that the current sample of high-z clumps with $\rm R_{eff}\leqslant20$ is very limited; new insight will come from samples observed with JWST, able to combine extreme spatial resolution with a better age and mass characterisation of the clumps \citep[see e.g.][]{claeyssens2022,vanzella2022_sunrise,vanzella2022_a2744}.
We also notice that at larger scales ($\rm R_{eff}>20$ pc, dashed violin distributions in Fig.~\ref{fig:sigmaM_literature}) only clumps from the $\rm z>1.5$ samples are observed to reach the most extreme densities ($\rm \Sigma_{M_\star}\gtrsim10^4\ M_\odot pc^{-2}$), again suggesting a redshift evolution even for large star-forming complexes.

As final remark, we remind that UV luminosities, and consequently $\rm SFR_{UV}$ values, are strongly affected by the age and extinction of the clumps; this could be the cause of the different strengths in the redshift evolution of $\rm \Sigma_{M_\star}$ and $\rm \Sigma_{SFR_{UV}}$.

\section{Conclusions}
\label{sec:conclusion}
We presented the analysis of a small sample of stellar clumps in three galaxies at redshift between 4 and 5, namely the lensed arcs beyond the galaxy clusters MS1358, RCS0224 and MACS0940. These galaxies were chosen to be among the most highly magnified systems hosting multiple stellar clumps in a redshift range currently under-represented in clump studies. 
Each galaxy is multiply-imaged; many of the images are amplified by factors $\rm \mu>5$ and reaching in some cases $\rm \mu>20$.
The clumps were studied using multi-band photometry from the HST, providing a maximum angular resolution, for the clump radii, of $\sim0.02''$; combined to the large amplifications of these systems, it allows the study of physical scales down to $\sim10$ pc, i.e. comparable to the sizes of individual stellar clusters. 

Clump populations in the three galaxies were extracted from a reference rest-frame UV filter (ACS-WFC-F775W for MS1358, ACS-WFC-F814W for RCS0224 and MACS0940); we further checked that our extraction did not miss clumps of different (redder) colors by testing the clump extraction also on the other available HST filters. Clump colors, in combination with the lens models, were used to recognise (and discard) foreground sources in the field; we find in total 10 unique clumps in MS1358, 3 in RCS0224 and 11 in MACS0940.

Clumps sizes and magnitudes were derived in the reference rest-frame UV filter for each of the galaxies; the other filters were used to fit a broad-band SEDs and derive the clump masses. 
Intrinsic sizes, magnitudes and masses were derived using the lens models; due to the large amplifications involved, the uncertainties associated to the amplification factors dominate these intrinsic quantities. 
The derived effective radii range from $\sim10$ to $\sim200$ pc; the smallest sizes are reached in RCS0224, the galaxy with the largest amplification, where we observe sources down to $\rm R_{eff}=6$ pc. UV magnitudes are also converted to $\rm SFR_{UV}$ following the conversion by \citet{kennicutt2012}; they range from $\sim10^{-2}$ to the most extreme values of $\rm 5\ M_\odot yr^{-1}$ in MS1358. 
The completeness limits of the samples, when converted into $\rm SFR_{UV}$ surface densities, $\rm \Sigma_{SFR}$, are typically $\rm \sim10\ M_\odot yr^{-1} kpc^{-2}$, and in all cases above $\rm 1\ M_\odot yr^{-1} kpc^{-2}$, i.e. above the typical value for clumps in local main sequence galaxies \citep{kennicutt2003,livermore2015}; we deduce that an eventual population of clumps with low surface brightnesses would be missed in this study.

By focusing individually on the galaxies of this study we find that: 
\begin{itemize}
    \item the morphology of MS1358 is dominated by two bright clumps, accounting for $20\%$ and $5\%$ of its rest-frame UV emission; when all the 10 observed clumps are considered together, this fraction raises to $\sim40\%$. The UV-derived SFR, $\rm SFR_{UV}$, of the brightest clumps are lower than the literature values derived from nebular emission \citep{swinbank2009}, suggesting either the presence of dust extinction or that the current star-formation episode is much younger than the assumption used in the \citet{kennicutt2012} FUV-to-SFR conversion. Clump masses range between $10^7$ and $10^9\ M_\odot$; they are consistent with the entire galaxy mass as derived by \citet{swinbank2009}. We deduce that clumps in MS1358 are major contributors to the recent star formation episode(s) in the galaxy and to the build-up of its mass.
    The two main clumps of MS1358 are the UV-brightest among the galaxies studied in the current work and among all the compact ($\rm R_{eff}<100$ pc) clumps known in literature. Despite most of the clumps in this galaxy have sizes in the range of $\rm 50-100$ pc, their mass surface densities are comparable (and in many cases higher) to the ones of the densest local stellar clusters.
    \item RCS0224 is characterised by 3 bright and compact clumps, accounting for $\sim45\%$ of the rest-frame UV emission. The galaxy also shows a larger region of diffuse UV emission, appearing as a very elongated arc devoid of clumps, but actively forming stars, as derived from nebular [\OII] emission by \citet{swinbank2007}; deeper observation would be needed to test the presence of low-surface brightness clumps along the arc. Despite RCS0224 has the shallowest data in the studied sample, its large magnification allows to reach very small intrinsic scales; the clumps sizes and masses, $\rm R_{eff}=6-25$ pc and $\rm M_\star=10^6-10^{7.5}\ M_\odot$ respectively, are close to the ones of the most massive stellar clusters in local galaxies. As is the case for clumps MS1358, also in RCS0224 the clump densities reach higher values than what typically observed in local samples.  
    \item Four bright clumps characterise the rest-frame UV morphology of the lensed arc MACS0940, accounting for $\sim40\%$ of the emission, with other 4 sources contributing to another $\sim10\%$. Their derived intrinsic sizes ($\rm R_{eff}=10-100$ pc) and masses ($\rm M_\star=10^6-10^8\ M_\odot$) suggest also for these clumps extreme stellar densities. 
\end{itemize}

Finally, we compare the SFR and stellar mass surface densities of the clumps to the ones of known samples in the literature. We find overall an increase of $\rm \Sigma_{SFR}$ with redshift, particularly when comparing clumps in local main sequence galaxies to their counterparts at cosmic noon $\rm z\sim1-3.5$; the evolution is less prominent at higher redshifts. A weaker evolution is suggested for $\rm \Sigma_{M_\star}$ (more evident for very compact sources, $\rm R_{eff}\leqslant20$ pc). We can interpret the evolution of both clump quantities in the context of the evolution of the properties of their host galaxies; the latter, at higher redshifts, are characterised by increasingly denser environments, which produce denser galactic (and sub-galactic) star-forming regions and, in turn, also denser stellar products. This interpretation is supported by the study of nearby starburst galaxies, whose clump properties resemble high-z samples more than the local average ones. 
With current data we do not find any discontinuity in the redshift evolution of SFR and $\rm M_\star$ surface densities when considering redshift earlier than cosmic noon ($\rm z>3.5$); this may imply that the clump formation conditions at $\rm z\sim5$ are not different compared to later times. This result is in line with recent ALMA findings suggesting the presence of turbulent disk galaxies even at $\rm z>5$ \citep{lelli2021,jones2021,rizzo2021,herrera2022,parlanti2023}. 

We remark that these redshift trends and relative interpretations are still tentative, for two main reasons. First, populations of clumps below $\sim100$ pc are still limited, leading to a small sample statistics when a binning in redshifts is considered. Second, the samples considered are in-homogeneous in how they have been extracted and analysed; this is true especially for what concerns SED-derived quantities, e.g. clump masses, ages and extinctions. On going observations with the JWST will go in the direction of tackling both problems by observing large galaxy samples and extending the rest-frame optical coverage also to high-redshift sources.

\section*{Acknowledgements}
This research made use of Photutils, an Astropy package for detection and photometry of astronomical sources \citep{bradley2020}. 
M.M. acknowledges the support of the Swedish Research Council, Vetenskapsrådet (internationell postdok grant 2019-00502). AA and AC acknowledge the support of the Swedish Research Council, Vetenskapsr{\aa}det (2021-05559)

\section*{Data Availability}
The HST data underlying this article are accessible from the Hubble Legacy Archive (HLA) at \url{https://hla.stsci.edu/} or through the MAST portal at \url{https://mast.stsci.edu/portal/Mashup/Clients/Mast/Portal.html}. Proposal IDs of the observations used: 9717 (PI: H. Ford) and 11591 (J. Kneib) for MS1358; 9135 (M. Gladders) and 14497 (R. Smit) for RCS0224; 11103 (H. Ebeling) and 15696 (D Carton) for MACS0940. The derived data generated in this research will be shared on reasonable request to the corresponding author.



\bibliographystyle{mnras}
\bibliography{clumps_at_z5} 




\appendix

\section{Best-fit models}
\label{sec:app:plots_bestfit}
\begin{figure*}
    \includegraphics[width=\textwidth]{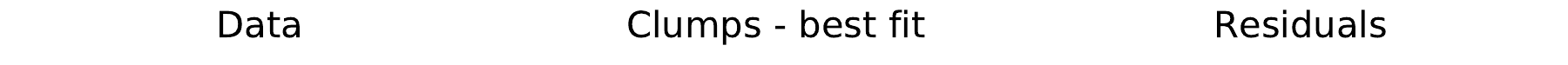} 
    \includegraphics[width=\textwidth]{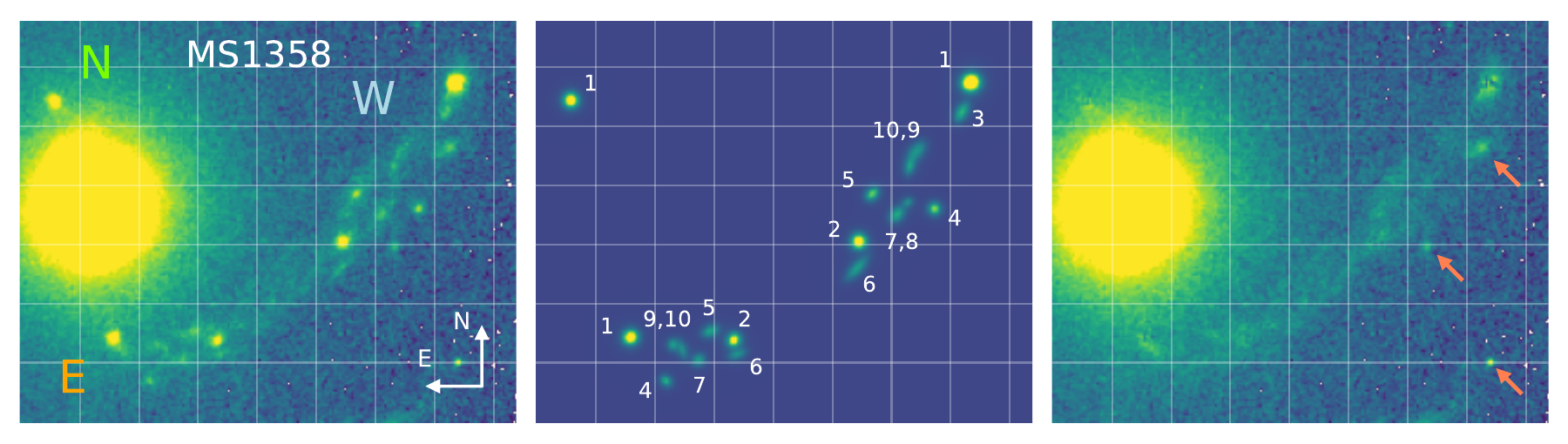}
    \includegraphics[width=\textwidth]{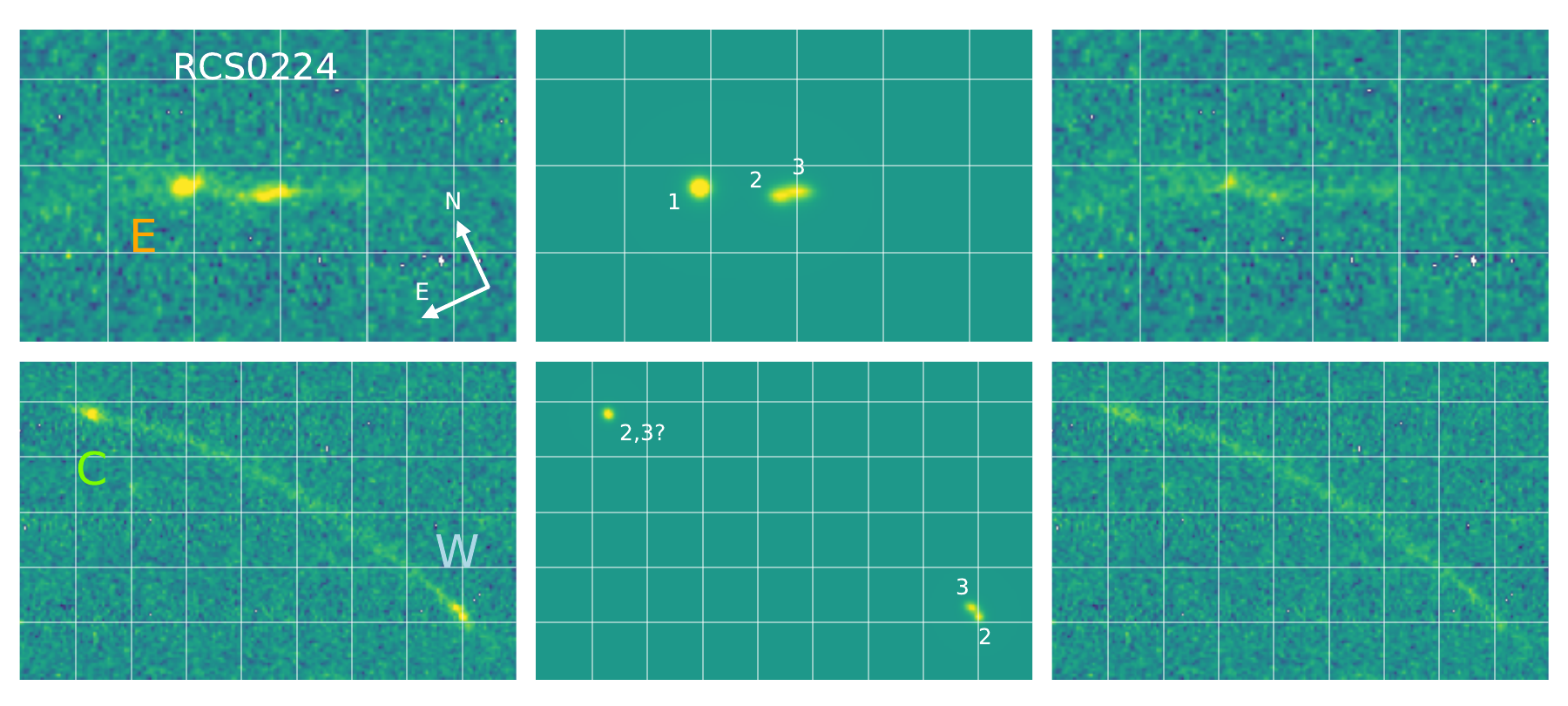}
    \includegraphics[width=\textwidth]{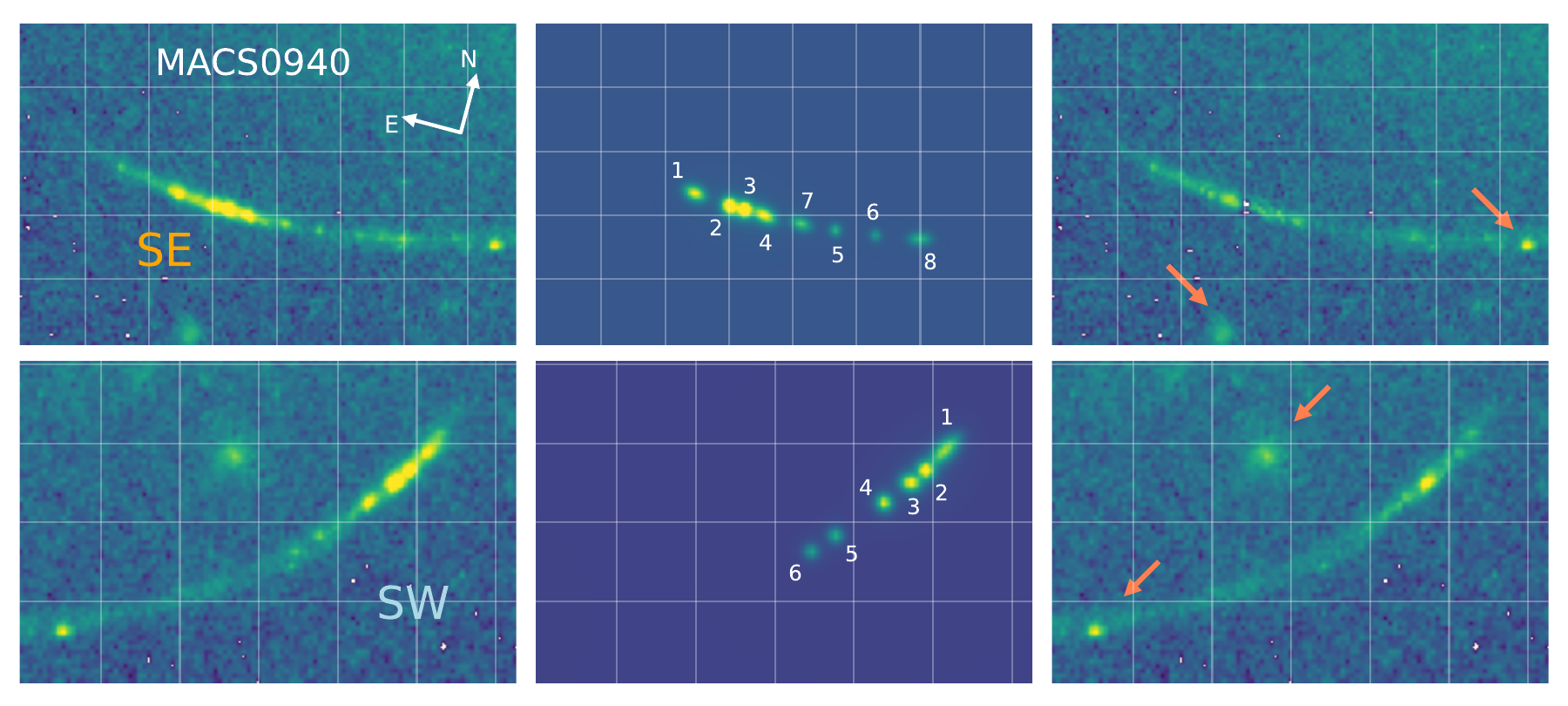}
    \caption{Data (left column), best-fit models (central column) and residuals (right column) for the main lensed arcs, in the reference filters of each galaxy. Clumps IDs are reported in the central panels. Foreground sources are marked with red arrows in the right-hand panels. Grids of $\rm 1\times1$ arcsec are plotted to facilitate the comparison between panels. For the position of the galaxy images within the cluster fields, see Fig.~\ref{fig:galaxies1},~\ref{fig:galaxies2} and~~\ref{fig:galaxies3}.}
    \label{fig:app:plots_bestfit}
\end{figure*}

We collect data, best–fit clump models and fit residuals in the reference filter of each galaxy in Fig.~\ref{fig:app:plots_bestfit}. 

\section{Additional plots}\label{sec:app:alternative_plots}
\begin{figure*}
    \includegraphics[width=0.56\textwidth]{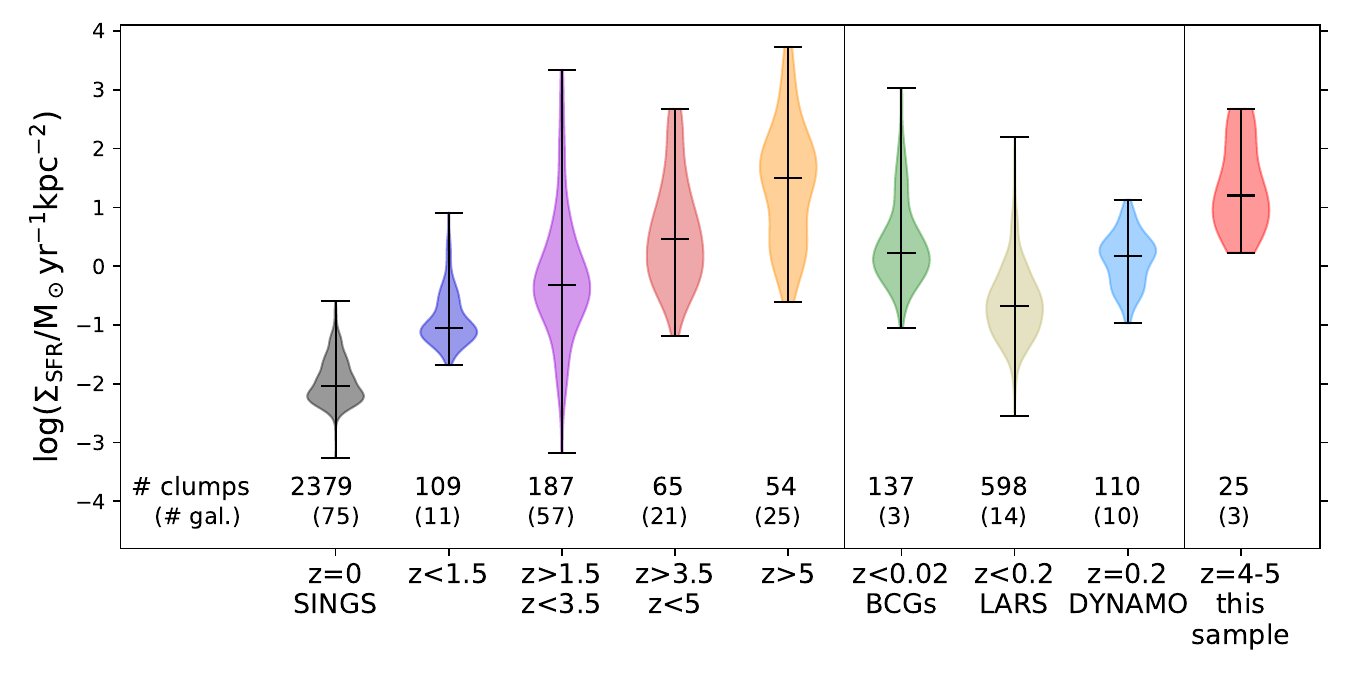} 
    \includegraphics[width=0.43\textwidth]{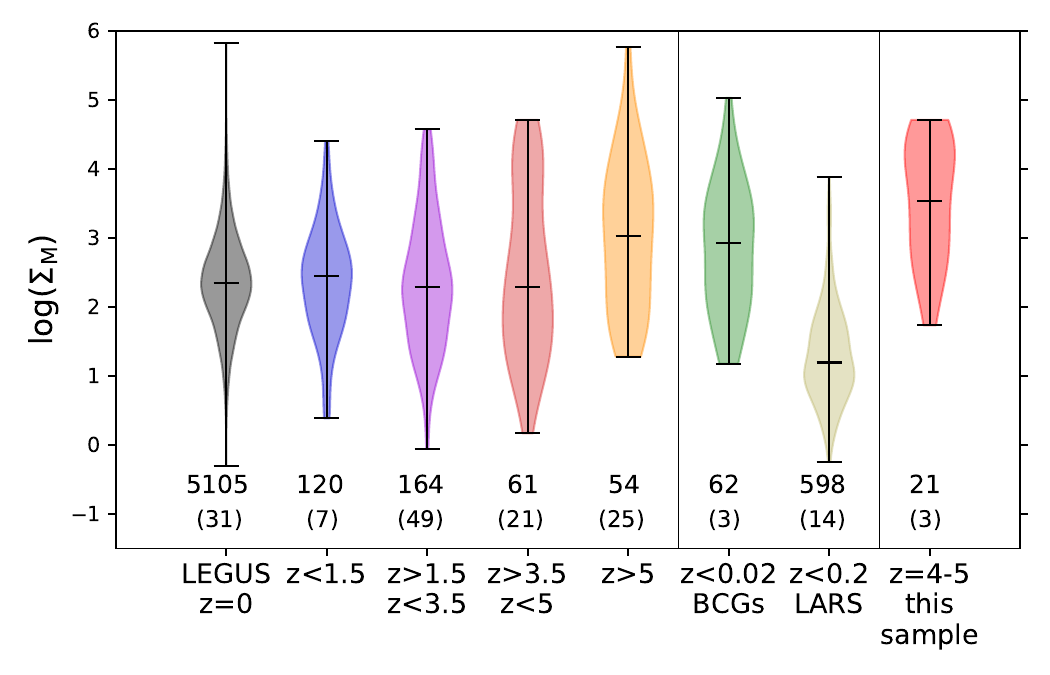} 
    \caption{(Left:) same as the bottom panel in Fig.~\ref{fig:sizelum_literature}, but including clumps of all sizes. (Right:) same as the right panel of Fig.~\ref{fig:sigmaM_literature} but including clumps of all sizes. The list and references of the samples used in both panels are given in the captions of Fig.~\ref{fig:sizelum_literature} and Fig.~\ref{fig:sigmaM_literature}.}
    \label{fig:app:sizelum_sigmaM_literature}
\end{figure*}

The distributions of \SigmaSFR\ across redshift bins, shown in Fig.~\ref{fig:sizelum_literature}, is expanded to include clumps of all sizes, loosening the constraint of $\rm R_{eff}\leq100$ pc, in Fig.~\ref{fig:app:sizelum_sigmaM_literature} (left panel). The main result reported in Section~\ref{sec:discussion_literature_sfr}, i.e. the overall redshift evolution of the \SigmaSFR\ distributions, up to redshifts $\rm z\sim3.5$, is valid also in this case; the main difference when comparing Fig.~\ref{fig:app:sizelum_sigmaM_literature} to Fig.~\ref{fig:sizelum_literature}, is the presence of more low-\SigmaSFR\ systems, reflected in longer tails at the bottom of the violin distributions. These low-density clumps are consistent with being `large' ($\rm R_{eff}>100$ pc) star forming regions.

In a similar way to what is done for the \SigmaSFR\ distributions, the right panel of Fig.~\ref{fig:app:sizelum_sigmaM_literature} shows violin distributions of \SigmaM, binned by redshift ranges and including clumps of all sizes. As already pointed out in Section~\ref{sec:discussion_literature_mass}, no clear redshift evolution for the mass surface density is observed for the overall sample; a tentative evolution can be deduced when only the very small clumps, with sizes consistent with individual stellar clusters ($\rm R_{eff}<20 pc$), are considered (right panel of Fig.~\ref{fig:sigmaM_literature}).

\bsp	
\label{lastpage}
\end{document}